\documentclass[aps,twocolumn,prd,dvipsnames,usenames,floatfix,superscriptaddress,nofootinbib,longbibliography,preprintnumbers]{revtex4-1}

\usepackage{amsmath,amssymb,bm}

\usepackage{fancyhdr} 
\usepackage{footnote}
\makesavenoteenv{tabular}
\makesavenoteenv{table}
\usepackage[utf8x]{inputenc}
\usepackage{units}
\usepackage[english]{babel}
\usepackage{xspace}
\usepackage{paralist}
\usepackage{amsmath} 
\usepackage{float}
\usepackage{amssymb} 
\usepackage{graphicx}
\usepackage{bm}
\usepackage{tabularx}
\usepackage{slashed}
\usepackage[final,textsize=footnotesize,color=ForestGreen!05]{todonotes}

\usepackage{subfig}
\usepackage[colorlinks=true,linktocpage=true,urlcolor=blue,citecolor=blue,linkcolor=blue]{hyperref}

\usepackage{dsfont}

\usepackage[justification=centerlast, format=plain, labelfont=bf,font=small]{caption}

\DeclareCaptionJustification{justified}{\leftskip=0pt \rightskip=0pt \parfillskip=0pt plus 1fil}

\makeatletter
\def\doauthor#1#2#3{%
  \ignorespaces#1\unskip
  \begingroup
   #3%
  \@if@empty{#2}{\@listcomma\endgroup{}{}}{\endgroup{\comma@space}{}\frontmatter@footnote{#2}}%
  \space \@listand
}%
\makeatother

\makeatletter
\def\@ssect@ltx#1#2#3#4#5#6[#7]#8{%
  \def\H@svsec{\phantomsection}%
  \@tempskipa #5\relax
  \@ifdim{\@tempskipa>\z@}{%
    \begingroup
      \interlinepenalty \@M
      #6{%
       \@ifundefined{@hangfroms@#1}{\@hang@froms}{\csname @hangfroms@#1\endcsname}%
       {\hskip#3\relax\H@svsec}{#8}%
      }%
      \@@par
    \endgroup
    \@ifundefined{#1smark}{\@gobble}{\csname #1smark\endcsname}{#7}%
  }{%
    \def\@svsechd{%
      #6{%
       \@ifundefined{@runin@tos@#1}{\@runin@tos}{\csname @runin@tos@#1\endcsname}%
       {\hskip#3\relax\H@svsec}{#8}%
      }%
      \@ifundefined{#1smark}{\@gobble}{\csname #1smark\endcsname}{#7}%
      \addcontentsline{toc}{#1}{\protect\numberline{}#8}%
    }%
  }%
  \@xsect{#5}%
}%
\makeatother

\makeatletter
\g@addto@macro\bfseries{\boldmath}
\makeatother

\begin{document}
\author{Maksym Ovchynnikov,}
\email{maksym.ovchynnikov@kit.edu}
\affiliation{Institut für Astroteilchen Physik, Karlsruher Institut für Technologie (KIT), Hermann-von-Helmholtz-Platz 1, 76344 Eggenstein-Leopoldshafen, Germany
}

\affiliation{Instituut-Lorentz for Theoretical Physics, Universiteit Leiden, Niels Bohrweg 2, 2333 CA Leiden, The Netherlands}
\author{Thomas Schwetz,}
\email{schwetz@kit.edu}
\affiliation{Institut für Astroteilchen Physik, Karlsruher Institut für Technologie (KIT), Hermann-von-Helmholtz-Platz 1, 76344 Eggenstein-Leopoldshafen, Germany
}
\author{Jing-Yu Zhu}
\email{jing-yu.zhu@kit.edu}
\affiliation{Institut für Astroteilchen Physik, Karlsruher Institut für Technologie (KIT), Hermann-von-Helmholtz-Platz 1, 76344 Eggenstein-Leopoldshafen, Germany
}
\affiliation{School of Physics and Astronomy and Tsung-Dao Lee Institute, Shanghai Jiao Tong University, 800 Dongchuan Rd, Shanghai 200240, China}

\title{
Dipole portal and neutrinophilic scalars at DUNE revisited:\\ 
The importance of the high-energy neutrino tail}

\begin{abstract}
\noindent We estimate the sensitivity of the DUNE experiment to new physics particles interacting with neutrinos, considering the dipole portal to heavy neutral leptons and a neutrinophilic scalar with lepton-number $2$ as examples. We demonstrate that neutrinos from the high-energy tail of the DUNE flux, with energies $E_{\nu}\gtrsim 5-10\text{ GeV}$, may significantly improve the sensitivity to these models, allowing to search for particles as heavy as $\simeq 10\text{ GeV}$. We also study the impact of the so-called tau-optimized neutrino beam configuration, which slightly improves sensitivity to the new physics models considered here. For both models, we consider new production channels (such as deep-inelastic scattering) and provide a detailed comparison of different signatures in the detector.
\end{abstract}

\maketitle

\tableofcontents

\section{Introduction}
In a broad class of Standard Model (SM) extensions, an interaction of SM neutrinos with new particles in the few GeV mass range is introduced.
Such models may be probed at neutrino experiments,  where a high-intensity neutrino beam is produced,
see e.g.~\cite{Batell:2022xau}. Examples of currently running and future neutrino beam experiments are 
T2K~\cite{T2K:2011qtm,T2K:2019bbb}, MiniBooNE~\cite{MiniBooNE:2001sbk}, MicroBooNE~\cite{MicroBooNE:2007ivj} and DUNE~\cite{DUNE:2020lwj}. Other promising facilities to look for such new particles are LHC-based experiments such as SND@LHC~\cite{SNDLHC:2022ihg},
FASER$\nu$~\cite{FASER:2020gpr}, or beam-dump experiments like SHADOWS~\cite{Baldini:2021hfw} and SHiP~\cite{SHiP:2015vad}.

The goal of this paper is to demonstrate the potential of the DUNE experiment to search for interactions of new physics particles with neutrinos. We do this by revising the sensitivity of DUNE to two example models -- a neutrinophilic scalar~\cite{Berryman:2018ogk,Kelly:2019wow} and the dipole portal to heavy neutral leptons (HNLs)~\cite{Magill:2018jla,Brdar:2020quo}. We improve on previous studies~\cite{Schwetz:2020xra,Atkinson:2021rnp,Berryman:2018ogk,Kelly:2019wow} in several aspects.
First, we include the high-energy neutrino tail in the calculation of the production rate, showing its importance to extend the reach in the mass of new physics particles, and we study the effect of the so-called tau-optimized beam configuration. In particular, we demonstrate that, depending on the model, DUNE may have sensitivity to new physics particles with masses up to $\mathcal{O}(10\text{ GeV})$ -- a few times larger than what it was obtained in the previous literature. Furthermore, for both models, we consider new production channels for the new particles. For the dipole portal, we discuss additional detection signatures; the ratio of the different signal types is a specific prediction of the dipole portal, allowing us to identify this model in case of detection. 

The paper is organized as follows. In Sec.~\ref{sec:neutrino-flux}, we discuss the flux of neutrinos at DUNE and stress the importance of its high-energy tail for the production of heavy particles. In Sec.~\ref{sec:dipole-portal}, we consider the dipole portal to HNLs, 
reestimate the DUNE sensitivity by considering new production channels such as deep inelastic scattering and new search signatures. In Sec.~\ref{sec:neutrinophilic-portal}, we discuss the neutrinophilic scalar portal, revising the sensitivity in a similar way. Finally, we conclude in Sec.~\ref{sec:conclusions}. Technical details, expressions for matrix elements and cross sections, as well as further discussion can be found in several appendices.

%%%%%%%%%%%%%%%%%%%%%%%%%%%%%%%%%%%%%%%%%%%%%%%%%%%%%%%%%%%%%%%%%%%%%
\section{High-energy neutrinos at DUNE}
\label{sec:neutrino-flux}

\subsection{The DUNE experiment}
\label{subsec:DUNE-experiment}
DUNE~\cite{DUNE:2020lwj} is the next-generation long-baseline neutrino oscillation experiment, primarily aiming at precisely measuring neutrino oscillation parameters, such as the unknown CP violation phase in the lepton sector. On the other hand, it is also a powerful tool to search for a variety of new physics beyond the SM.
By using a 120 GeV proton beam onto a graphite target with a beam power of 1.2 MW, this experiment can provide 1.1 $\times 10^{21}$ PoT/year, generating a large flux of light mesons such as $\pi^{\pm},K^{\pm,0}$. A fraction of these mesons then would decay inside a 194 meter long decay pipe, producing neutrinos. As a result, DUNE provides the world's most intense neutrino beam with a wide range of neutrino energies peaking at about 2.5 GeV. The interactions of neutrinos are supposed to be studied in two detectors -- the near detector (ND) and the far detector (FD). The former is designed to include a 67.2 ton Liquid Argon Time Projection Chamber (LArTPC), a magnetized gaseous argon time projection chamber, and a large, magnetized monitor, which will be located 574 m downstream of the neutrino target following the decay pipe. The FD will be equipped with four 10-kton LArTPC modules at a distance of 1285 km from the target; one can refer to Fig~1.4 in Ref.~\cite{DUNE:2020txw} for the layout of the modules. The useful parameters of ND and FD are summarized in Table~\ref{tab:DUNE-parameters}.

\begin{table}[t]
    \centering
    \begin{tabular}{|c|c|c|c|c|}
    \hline  Detector &  $S_{\text{transverse}}$ &$L_{\text{det}}$ &   $L_{\text{to det}}$ & $n_{\text{nucl}}$\\ \hline
     ND    & $2\times 6\text{ m}^{2}$ &  4 m &574 m  & $8.4\times 10^{29}\text{ m}^{-3}$  \\ \hline FD &  $12\times 14\text{ m}^{2}$ & $58.2$ m &  $1285$ km & $8.4\times 10^{29}\text{ m}^{-3}$ \\ \hline
    \end{tabular}
    \caption{Parameters of Near Detector and individual Far Detector module at DUNE. The numbers are taken from~\cite{DUNE:2020txw,DUNE:2020fgq}, where $S_{\rm transverse}$, $L_{\rm det}$, $L_{\rm to ~det}$, and $n_{\text{nucl}}$ stand for the fiducial cross-sectional area of the detector, the fiducial length of the detector, the distance to the proton collision point, and the nucleon number density, respectively. }
    \label{tab:DUNE-parameters}
\end{table}

In order to maximize the flux of neutrinos/antineutrinos in the direction of the DUNE detectors, the charged particles need to be collimated with respect to the beam axis by a magnetic horn system. In dependence on the operating mode (the one for neutrino or antineutrino), correspondingly positively or negatively charged particles would be collimated. The flux in antineutrino mode is very similar to the one in neutrino mode (see e.g., Fig.~5.4 in~\cite{DUNE:2020lwj}), and we expect similar sensitivities to the new physics models discussed below for neutrino and antineutrino beam modes. To be specific, in this work, we will only consider the neutrino mode.

In addition, there are two horn configurations considered by the DUNE collaboration: the CP-optimized configuration, which maximizes the flux at $E_{\nu_{\tau}}<5\text{ GeV}$ important to study CP violations in neutrino oscillations, and the tau-optimized configuration, for which the flux of $\tau$ neutrinos at the FD with $E_{\nu}<5\text{ GeV}$ is somewhat lower, but the higher-energy flux between 5 and 10~GeV gets significantly increased, which would result in an order of magnitude higher number of neutrino deep-inelastic scattering (DIS) events~\cite{DUNE:2020lwj}.

%%%%%%%%%%%%%%%%%%%%%%%%%%%%%%%%%%%%%%%%%%%%%%%%%%%%%%%%%%%
\subsection{Neutrino flux at DUNE}

\begin{figure}[t]
    \centering    \includegraphics[width=0.45\textwidth]{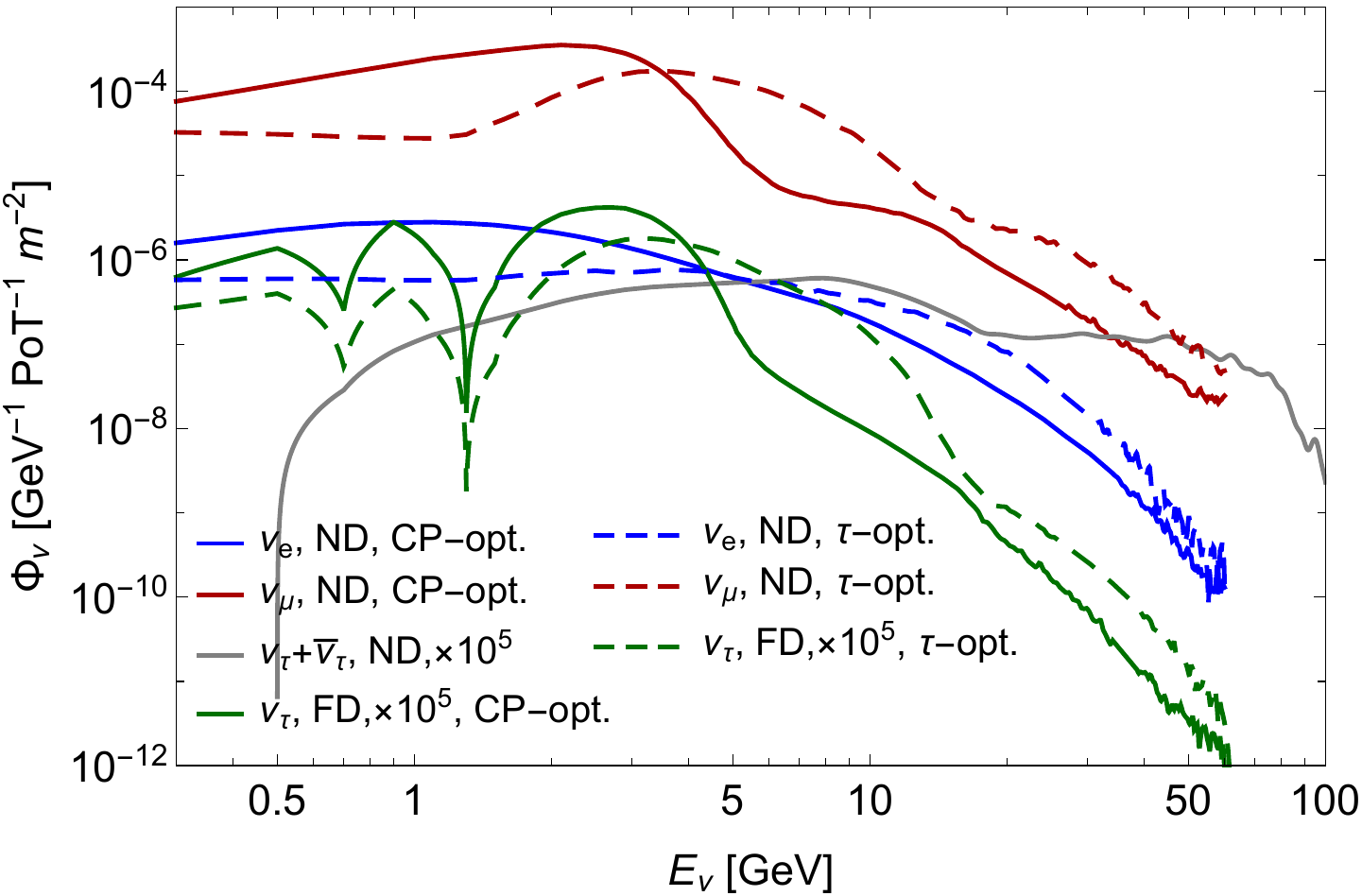}
    \caption{Neutrino fluxes as defined in Eq.~\eqref{eq:fluxes} of $\nu_{e,\mu,\tau}$ at the DUNE ND and the flux of $\nu_{\tau}$ at the DUNE FD (due to $\nu_\mu\to\nu_\tau$ oscillations), assuming neutrino operating mode of the focusing horns. $\nu_\tau$ fluxes are magnified by a factor $10^5$ for better visibility.
    The results for two beam configurations are shown: the CP-optimized (solid), and the tau-optimized (dashed). Note that the $\nu_{\tau}$ flux at the ND does not depend on the horn configuration, since it originates from decays of promptly decaying $D_{s}$ mesons and $\tau$ leptons.}
    \label{fig:neutrino-fluxes}
\end{figure}

We define the neutrino flux at a given detector as
\begin{equation}
\Phi_{\nu} \equiv \frac{1}{N_{\text{PoT}}\cdot S_{\text{transverse}}}\frac{dN_{\nu}}{dE_{\nu}},
\label{eq:fluxes}
\end{equation}
where $N_{\text{PoT}}$ is the number of protons on target, $S_{\text{transverse}}$ is the transverse area of the detector (Table~\ref{tab:DUNE-parameters}), $dN_{\nu}/dE_{\nu}$ is the differential distribution of neutrinos traveling in the direction of the detector.
The fluxes at ND and FD for various neutrino flavors, assuming the neutrino operating mode, are shown in Fig.~\ref{fig:neutrino-fluxes}. Below, we briefly discuss their main characteristics.

The electron and muon neutrinos at DUNE are produced mainly by decays of light long-lived mesons such as $\pi^{\pm}, K^{\pm}, K^{0}_{L}$, and muons. The low-energy part of the spectra of $\nu_{\mu}$ ($E_{\nu_{\mu}}\lesssim 6\text{ GeV}$) and $\nu_{e}$ ($E_{\nu_{e}}\lesssim 10\text{ GeV}$) originates from decays 
\begin{equation}
\pi^{+}\to \nu_{\mu}+\mu^{+}, \quad \mu^{+} \to \nu_{e}+\bar{\nu}_{\mu}+e^{+}
\end{equation}
correspondingly. The high-energy tail comes from decays of kaons $K^{+}, K^{0}_{L}$. The relative suppression $\nu_{e}/\nu_{\mu}$ comes from the fact that muons are long-lived, $\tau_{\mu}/\tau_{\pi/K}\propto 10^{2}$, and only a small fraction of them, $\simeq 10^{-2}$, decays inside the decay pipe before being scattered/absorbed in the material after the decay pipe. To obtain the fluxes of these neutrinos, we use the publicly available results of the detailed GEANT4~\cite{GEANT4:2002zbu, Allison:2016lfl, Allison:2006ve} based simulation (G4LBNF) of the LBNF beamline developed by the DUNE collaboration~\cite{DUNE:2020lwj, DUNEfluxes}. Technical details are given in Appendix~\ref{app:neutrino-meson-flux}. 

When considering the DUNE sensitivity to new physics interacting with $\nu_{e,\mu}$, we will only consider the ND. The reason is that the fluxes of $\nu_{e/\mu}$ at the FD are much smaller than at the ND due to the much smaller angular coverage of the FD. To argue this point, let us compare the products of the flux times the detector volume $V_{\text{det}}$, $\Phi \times V_{\text{det}}$, at the ND and FD. Using that $V_{\text{FD}}/V_{\text{ND}}\approx 800$ if assuming four FD module (Table~\ref{tab:DUNE-parameters}), and neglecting the oscillations for the moment, one has
\begin{equation}
    \frac{\Phi_{\nu_{e/\mu}}^{\text{FD}}V_{\text{FD}}}{\Phi_{\nu_{e/\mu}}^{\text{ND}}V_{\text{ND}}} \sim 800\times \left(\frac{L_{\text{to ND}}}{L_{\text{to FD}}}\right)^{2} = 1.6\cdot 10^{-4},
    \label{eq:ND-to-FD-electron-muon}
\end{equation}
where $(L_{\text{to ND}}/L_{\text{to FD}})^{2}\approx 2\cdot 10^{-7}$ is the solid angle suppression (see Table~\ref{tab:DUNE-parameters}). Thus, the FD is less relevant for searching for hypothetical BSM particles interacting with these flavors. 

Let us now discuss $\tau$ neutrinos. At the DUNE ND, their main production channels are the decays 
\begin{equation}
D_{s}^{+}\to \tau^{+} + \nu_{\tau}, \quad \tau^{+} \to \bar\nu_{\tau}+X
\label{eq:tau-neutrinos-production}
\end{equation}
and their charge conjugated channels, where $X$ denotes lepton or hadron final states. Since $D_{s}$ and $\tau$ decay promptly, their distribution (and hence the flux of $\tau$ neutrinos and antineutrinos at ND) is not affected by the horn configuration. However, the flux of these $D_s/\tau$-originated $\nu_{\tau}$s or $\bar\nu_{\tau}$s at the FD would be too small compared to the flux of $\nu_{\tau}$ originated from the oscillations $\nu_{\mu}\to \nu_{\tau}$, and therefore the former can be safely ignored. Only the latter ($\nu_\tau$ from oscillations) would be present at the FD when we calculate the sensitivities of neutrino mode to new physics.

The DUNE simulations did not include $D_s$ mesons and $\tau$ leptons. To generate the $\tau$ neutrino flux, we have used the spectrum of $D_{s}$ mesons from~\cite{Krasnov:2019kdc}, then simulated the decay chain~\eqref{eq:tau-neutrinos-production} (see the Appendix~\ref{app:neutrino-meson-flux}), and selected the $\tau$ neutrinos that point to the ND.

The relative flux suppression $\nu_{\tau}/\nu_{\mu}$ at the ND parametrically behaves as
\begin{equation}
\frac{\Phi^{\text{ND}}_{\nu_{\tau}+ \bar\nu_{\tau}}}{\Phi^{\text{ND}}_{\nu_{\mu}}} \simeq \frac{P_{pp\to D_{s}}\cdot \text{Br}(D_{s}\to \tau)}{P_{pp\to \pi^{+}}}p_{E_{\nu}} \simeq  3\cdot 10^{-8}\times p_{E_{\nu}} \,.
    \label{eq:flux-nu-tau-ND}
\end{equation}
Here, $P_{pp\to D_{s}}\simeq 4\cdot 10^{-6}$, $P_{pp\to \pi^{+}}\approx 6.3$ are multiplicities (the number of decayed mesons per PoT), taken from~\cite{Coloma:2020lgy}, $\text{Br}(D_{s}\to \tau)\approx 0.055$~\cite{ParticleDataGroup:2020ssz}, and $p_{E_{\nu}}$ is a factor depending on the neutrino energy,
varying from $\mathcal{O}(1)$ at $E_{\nu}\lesssim 3\text{ GeV}$ to $\simeq 10^{3}$ for $E_{\nu}\simeq 50\text{ GeV}$. The reason of this behavior is that $\tau$ leptons (and hence $\tau$ neutrinos) have much larger mean energy than pions and kaons (and hence $\nu_{e/\mu}$) decaying inside the decay pipe.

Apart from the production in decays of mesons and $\tau$ leptons, $\nu_{\tau}$ may also be produced via oscillations $\nu_{\mu}\to \nu_{\tau}$. This channel is not relevant for the ND, since the typical neutrino oscillation length is much larger than the distance from the target to ND. However, it becomes the main contribution to the $\nu_{\tau}$ flux at the FD (see Sec.~\ref{sec:neutrino-flux}). To obtain the oscillated flux at the FD, we have first extracted the flux of $\nu_{\mu}$ at the FD, and then convoluted it with the oscillation probability
\begin{equation}
    P_{\nu_{\mu}\to \nu_{\tau}} \approx 0.943\times \sin^{2}\left( \frac{\Delta m^{2}L}{4E_{\nu}}\right)
\end{equation}
assuming $\quad \Delta m^{2} = 2.523\times 10^{-3}\text{ eV}^{2}$ and $L \approx 1300\text{ km}$ for all neutrinos. The resulting flux is
\begin{equation}
    \Phi_{\nu_{\tau}}^{\text{FD}} = \Phi_{\nu_{\mu}}^{\text{FD}} \times P_{\text{osc}} \simeq \Phi_{\nu_{\mu}}^{\text{ND}}\times \left(\frac{L_{\text{to ND}}}{L_{\text{to FD}}}\right)^{2} \times 
    P_{\nu_{\mu} \to \nu_{\tau}},
    \label{eq:flux-nu-tau-FD}
\end{equation}
where $P_{\nu_{\mu} \to 
\nu_{\tau}}$ is ${\cal O}(1)$ at the FD. Since the oscillation probabilities $\nu_{\mu}\to \nu_{\tau}$ and $\bar{\nu}_{\mu}\to\bar{\nu}_{\tau}$ are the same up to small CP-violating effects, independently on the oscillating neutrino energy, the property of the relative suppression of the antineutrino to neutrino fluxes in the neutrino mode translates to the fluxes of $\tau$ neutrinos and antineutrinos at the FD.

Therefore, the $\nu_{\tau}$ fluxes at the ND and FD have a different origin. As a result, the conclusion that the FD is not relevant for the DUNE sensitivities to new physics becomes invalid in the case of interactions with $\tau$ flavor. To illustrate this, let us again compare the products of the flux times the detector volume. For the CP-optimized horn configuration, using Eqs.~\eqref{eq:flux-nu-tau-ND},~\eqref{eq:flux-nu-tau-FD}, one has
\begin{equation}
    \frac{\Phi^{\text{FD}}_{\nu_{\tau}}\times V_{\text{FD}}}{\Phi^{\text{ND}}_{\nu_{\tau}}\times V_{\text{ND}}}\simeq 1\cdot 10^{3} \frac{P_{\text{osc}}(E_{\nu})}{p_{E_{\nu}}} \;.
    \label{eq:ratio-nu-tau-fd-nd}
\end{equation}
Considering the FD and $E_{\nu}\lesssim 5$ GeV, both $P_{\rm osc}$ and $p_{E_{\nu}}$ can be of ${\cal O}(1)$, leading to the ratio in Eq.~\eqref{eq:ratio-nu-tau-fd-nd} being very large. Hence, the FD may provide better sensitivities to new physics coupling to $\nu_\tau$~\cite{Schwetz:2020xra}.
However, since $P_{\text{osc}}(E_{\nu}) \sim E_{\nu}^{-2}$ at energies $E_{\nu}\gtrsim 10\text{ GeV}$, together with large $p_{E_{\nu}}$ they cause the suppression of this ratio at large energies. As a result, with the increase of the neutrino energy, the ratio~\eqref{eq:ratio-nu-tau-fd-nd}, being $\gg 1$ at $E_{\nu} = \mathcal{O}(1\text{ GeV})$, quickly drops and becomes $\mathcal{O}(1)$ at $E_{\nu} \simeq 15\text{ GeV}$. 

The situation is somewhat different for the tau-optimized horn configuration, for which tau neutrinos at the FD are more energetic on average (see Fig.~\ref{fig:neutrino-fluxes}), although qualitatively the conclusions do not change. Therefore, both the ND and FD may be important for searching for new particles coupling to $\nu_{\tau}$, depending on their mass.

\begin{table}[t]
    \centering
    \begin{tabular}{|c|c|c|c|c|}
    \hline Detector & Horn conf. & $N_{\nu_{e}}$ & $N_{\nu_{\mu}}$ & $N_{\nu_{\tau}}$  \\ \hline
    ND & CP-optimized  & $1.3\cdot 10^{-4}$  & $1.2\cdot 10^{-2}$ & $7.5\cdot 10^{-10}$ \\ \hline 
    ND & tau-optimized  & $1.0\cdot 10^{-4}$  & $1.4\cdot 10^{-2}$ & $7.5\cdot 10^{-10}$ \\ \hline
    FD & CP-optimized  & $1.8\cdot 10^{-9}$ & $8.9\cdot 10^{-9}$ & $1.5\cdot 10^{-8}$ \\ \hline 
    FD & tau-optimized  & $1.05\cdot 10^{-9}$ & $1.4\cdot 10^{-8}$ & $1.1\cdot 10^{-8}$ \\ \hline 
    \end{tabular}
    \caption{Numbers of neutrinos per PoT within the angular acceptance of the DUNE detectors (ND or FD), assuming the parameters of the detectors given by Table~\ref{tab:DUNE-parameters}. Neutrino oscillations are included. Two horn configurations are assumed: CP-optimized and tau-optimized. For FD, we report numbers corresponding to one module.}
    \label{tab:DUNE-neutrino-numbers}
 % Table II
\end{table}

The total numbers of neutrinos traveling in the direction of the DUNE detectors (with the parameters given in Table~\ref{tab:DUNE-parameters}) per PoT are given in Table~\ref{tab:DUNE-neutrino-numbers}.

%%%%%%%%%%%%%%%%%%%%%%%%%%%%%%%%%%%%%%%%%%%%%%%%%%%%%%%%%%%%%%%
\subsection{High-energy neutrinos and production of new physics particles}
\label{sec:neutrino-thresholds}

In Refs.~\cite{Schwetz:2020xra,Kelly:2019wow}, which have studied the sensitivity of DUNE to new physics particles produced in scatterings of electron and muon neutrinos, artificial cuts on the neutrino spectrum have been imposed: $E_{\nu}<6\text{ GeV}$ and $E_{\nu}<10\text{ GeV}$ correspondingly. This is reasonable if one studies the production of particles $Y$ with mass $m_{Y}\ll E_{\nu,\text{max}}$. Indeed, first, the production of such particles does not require large energies. Second, only a tiny fraction of $\nu_{e},\nu_{\mu}$ have energies $E_{\nu}>5-10\text{ GeV}$, see Fig.~\ref{fig:neutrino-fluxes}. 

However, as we show below, these cuts can significantly underestimate the maximal mass of new physics particles that may be searched for at DUNE. Namely, the sensitivity of DUNE estimated in~\cite{Schwetz:2020xra,Kelly:2019wow} rapidly drops at masses $m_{Y} = 2.5-3\text{ GeV}$, which is directly related to the cuts. To understand this, let us look closer at the relevant $Y$ production processes (here without specifying the model details). Correspondingly, they are\footnote{Apart from the scatterings off nucleons,~\cite{Schwetz:2020xra} considered scatterings off electrons and nuclei, but the 
former requires much larger neutrino energies to produce a particle with the given mass, while the latter is suppressed due to nuclear form factors for large masses $m\gg r_{\text{nuclear}}^{-1}$.} 
\begin{equation}
    \nu + p\to Y+p, \quad \nu + p \to Y+\mu + n,
    \label{eq:production-process-example}
\end{equation}
where $n,p$ are nucleons, and in the second process a threshold for the invisible transverse momentum of $p_{T,Y}>p_{T,\text{min}} = 0.5\text{ GeV}$ is required to suppress backgrounds (we will discuss this in more detail in Sec.~\ref{sec:neutrinophilic-portal}).

\begin{figure}[t]
    \centering
    \includegraphics[width=0.45\textwidth]{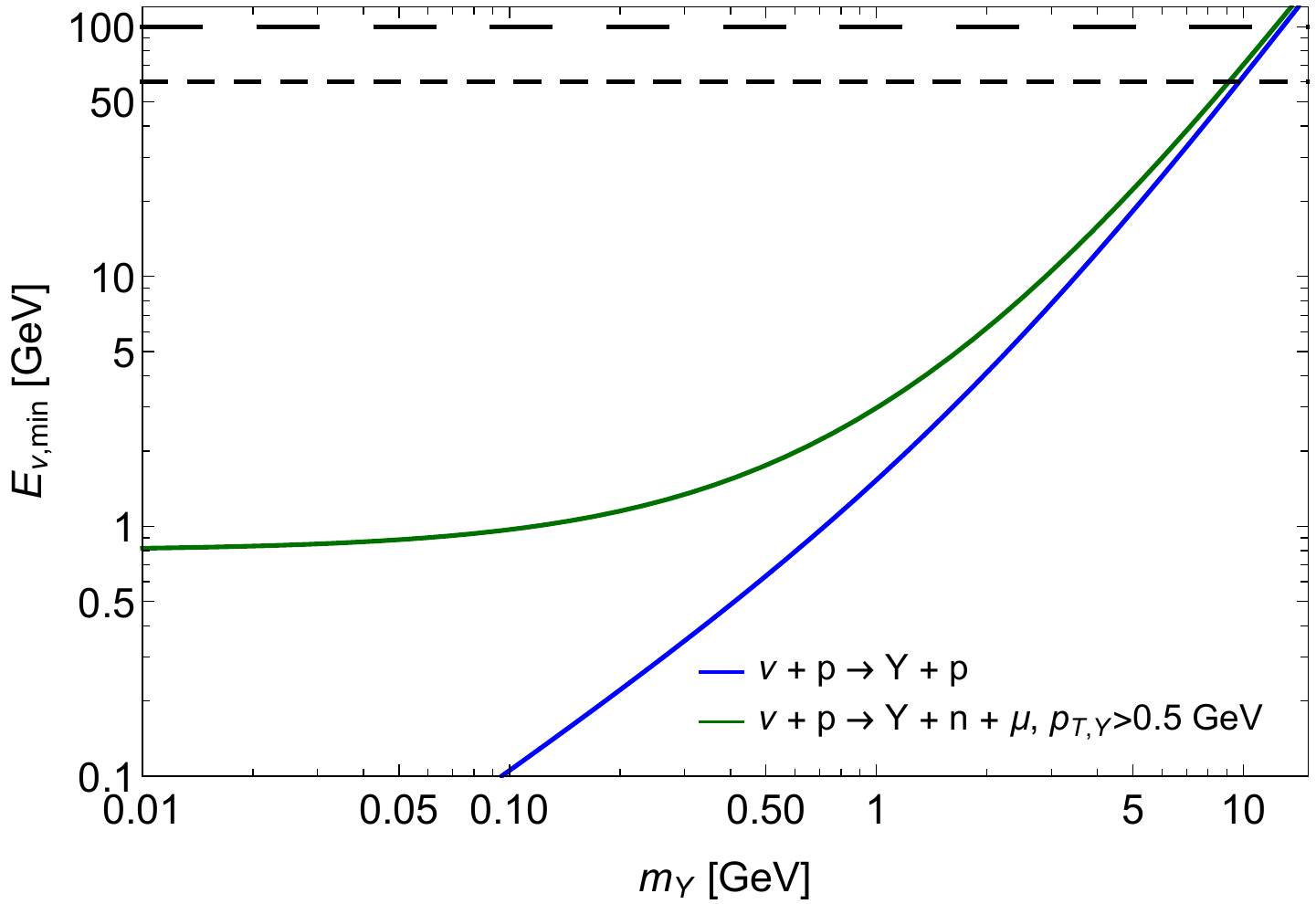}
    \caption{Dependence of the minimal neutrino energies $E_{\nu,\text{min}}^{(1),(2)}$, Eq.~\eqref{eq:min-energy}, required to produce a particle $Y$ in the scattering processes in Eq.~\eqref{eq:production-process-example}, on the $Y$ mass. The short- and long-dashed black lines denote correspondingly the maximal energies of $\nu_{e,\mu}$ and $\nu_{\tau}$ obtained in the simulation (see Fig.~\ref{fig:neutrino-fluxes}).}
    \label{fig:Enumin}
\end{figure}

The minimal energy $E_{\nu,\text{min}}$ of the neutrino required to produce the particle $Y$ in these processes is, correspondingly,
\begin{align}
    E_{\nu,\text{min}}^{(1)} &=  \frac{2m_{p}m_{Y}+m_{Y}^{2}}{2m_{p}}\,,\\ 
E_{\nu,\text{min}}^{(2)} &\simeq \nonumber\\ &\frac{\left[m_n+m_\mu+\sqrt{m_Y^2 + p_{T,{\rm min}}^2} +\frac{p_{T,{\rm min}}^2} {2 (m_n + m_\mu)}\right]^2
-m_{p}^{2}}{2m_{p}} \,.
    \label{eq:min-energy}
\end{align}

The behavior of $E^{(1),(2)}_{\nu,\text{min}}$ as a function of $m_Y$ is shown in Fig.~\ref{fig:Enumin}.
We see from the figure that at DUNE it actually may be possible to produce much heavier particles -- up to $m_{Y}\simeq 8 \text{ GeV}$ (the production from $\nu_{e/\mu}$), or to $m_{Y}\simeq 10\text{ GeV}$ (from $\nu_{\tau}$) if neutrinos with energies up to 100~GeV are taken into account, offering a potential trade-off to the reduced flux at high energies.

In the next two sections, we will study how the DUNE sensitivity to the mentioned models extends in detail. We will consider both ND/FD CP-optimized and tau-optimized horn configurations.

%%%%%%%%%%%%%%%%%%%%%%%%%%%%%%%%%%%%%%%%%%%%%%%%%%%%%%%%%%%%%%%%
\section{Neutrino dipole portal}\label{sec:dipole-portal}
%%%%%%%%%%%%%%%%%%%%%%%%%%%%%%%%%%%%%%%%%%%%%%%%%%%%%%%%%%%%%%%%

The effective Lagrangian of the neutrino dipole portal below the electro-weak (EW) scale is~\cite{Magill:2018jla}
\begin{equation}
    \mathcal{L}_{\text{dipole}} = d_{\alpha} \bar{N}\sigma_{\mu\nu}P_{L}\nu_{\alpha}F^{\mu\nu}+\text{h.c.}, 
    \label{eq:dipole-portal}
\end{equation}
where $\nu_{\alpha}$ is the SM neutrino of flavor $\alpha=e,\mu,\tau$, $\sigma_{\mu\nu} = \frac{i}{2}[\gamma_{\mu},\gamma_{\nu}]$, $P_{L} = (1-\gamma_{5})/2$, $F_{\mu\nu} = \partial_{\mu}A_{\nu}-\partial_{\nu}A_{\mu}$ is the electromagnetic field strength tensor, and $N$ is a Heavy Neutral Lepton (HNL). 

Note that this Lagrangian is not gauge-invariant and therefore valid only at energies below the EW scale, above which we need to consider UV-completions of the operator in Eq.~\eqref{eq:dipole-portal}. Here we remain agnostic about the UV origin of this new interaction and study its phenomenological implications at energies below the EW scale.

Motivated by the unsolved MiniBooNE~\cite{MiniBooNE:2018esg}, ANITA~\cite{ANITA:2016vrp,ANITA:2018sgj}, and muon g-2 anomalies~\cite{Muong-2:2021ojo,Babu:2021jnu}, the dipole portal provides another way to test the existence of HNLs and has attracted a lot of attention recently~\cite{Schwetz:2020xra,Gninenko:2009ks,Coloma:2017ppo,Shoemaker:2020kji,Plestid:2020vqf,Jodlowski:2020vhr,Atkinson:2021rnp,Dasgupta:2021fpn,Dasgupta:2021fpn,Ismail:2021dyp,Miranda:2021kre,Miranda:2021kre,Bolton:2021pey, Arguelles:2021dqn,Mathur:2021trm,Li:2022bqr,Zhang:2022spf,Huang:2022pce,Gustafson:2022rsz,Kamp:2022bpt,Abdullahi:2022cdw,Delgado:2022fea}. 
Bounds on $d_\alpha$ come from various laboratory, astrophysical and cosmological observations. Laboratory constraints come from neutrino oscillation experiments, dark matter detectors, and the observation of high-energy neutrinos in neutrino telescope by studying coherent elastic neutrino-nucleus scattering, elastic neutrino-electron scattering, deep inelastic interactions, etc. Astrophysical constraints on $d_{\alpha}$ arise from supernova bursts, Big Bang Nucleosynthesis, or Cosmic Microwave Background. We refer to Refs.~\cite{Magill:2018jla, Brdar:2020quo,Kamp:2022bpt} for a compilation of various constraints.

%%%%%%%%%%%%%%%%%%%%%%%%%%%%%%%%%%%%%%%%%%%%
\subsection{Phenomenology at DUNE}
\label{sec:dipole-portal-DUNE}

\begin{figure*}[!t]
            \centering
           \includegraphics[width=\textwidth]{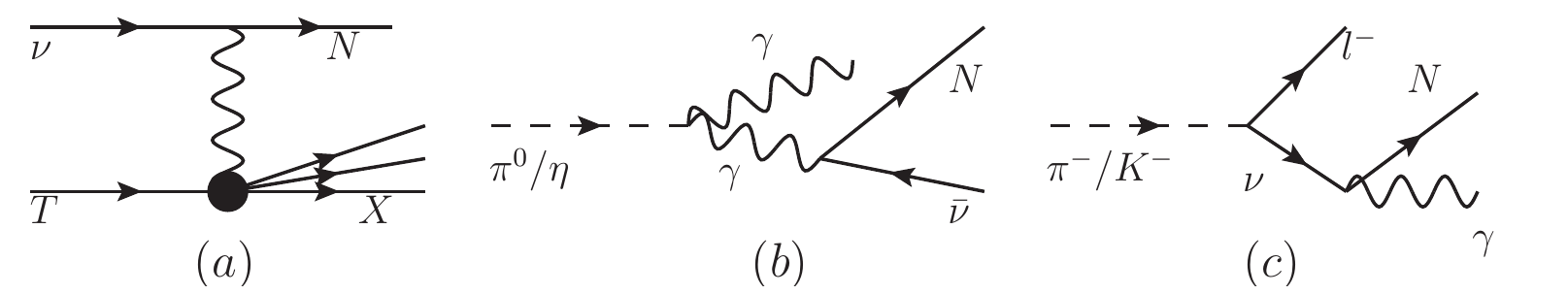}
            \caption{Production channels of HNLs via the dipole portal in  Eq.~\eqref{eq:dipole-portal} through (a): neutrino up-scattering, (b): promptly decaying mesons $\pi^{0}$ or $\eta$, and (c): decays of long-lived mesons $\pi^{-}$ or $K^{-}$. For the neutrino up-scattering process, $T$ and $X$ denote electron, nucleon, nucleus, or an arbitrary hadronic state.}
           \label{fig:dipole-portal-production-diagrams}
\end{figure*}    

The DUNE sensitivity to HNLs with a dipole portal interaction has been studied previously in Refs.~\cite{Schwetz:2020xra, Atkinson:2021rnp}. The HNL production mechanism studied there was mainly quasi-elastic (QE) neutrino up-scattering~\cite{Magill:2018jla},
\begin{equation}
    \nu_{\alpha}+T \to N+T, 
\end{equation}
where $T = e,n/p,\text{Ar}$ or atomic nuclei in the crust along the trajectory of the neutrino beam. 

Here, we will include also the deep-inelastic (DIS) contribution, $\nu_{\alpha}+p/n \to N+X$, where $X$ is an arbitrary hadronic state, 
see Fig.~\ref{fig:dipole-portal-production-diagrams} (a). Furthermore, in addition to the neutrino up-scattering, we consider the production of HNLs by decays of short-lived and long-lived mesons~\cite{Magill:2018jla}, $\pi^{\pm,0},K^{\pm,0},\eta,\rho^{0}$, see Fig.~\ref{fig:dipole-portal-production-diagrams} (b), (c) for two example diagrams. A detailed discussion of the relevant cross sections and production rates is given in Appendix~\ref{app:dipole-portal-production}. The HNL production may occur either inside the detector (neutrino up-scattering) or outside it (both neutrino up-scattering and meson decays). In the latter case, HNLs need to reach the detector in order to decay inside it and hence be detected. As discussed in detail in Appendix~\ref{app:dipole-portal-production-comparison}, in most cases the production from meson decays plays only a sub-leading role and the main production channel is either inside or outside up-scattering. The only exception is HNL production via $d_\tau$ at the ND. 

The main HNL decay channels are 
\begin{equation}
    N\to \gamma + \nu_\alpha, \quad N \to l^{+}+l^{-}+\nu_\alpha \,,
    \label{eq:dipole-production-processes}
\end{equation}
with $l = e,\mu$. The dominant channel is the mono-photon channel, with the decay width being 
\begin{equation}
   \Gamma_{N\to \nu_{\alpha}\gamma} =  \frac{|d_{\alpha}|^{2}m_{N}^{3}}{4\pi}
   \label{eq:dipole-decay-width}
\end{equation}
for Dirac neutrinos.

\begin{figure}[t]
    \centering
    \includegraphics[width=0.45\textwidth]{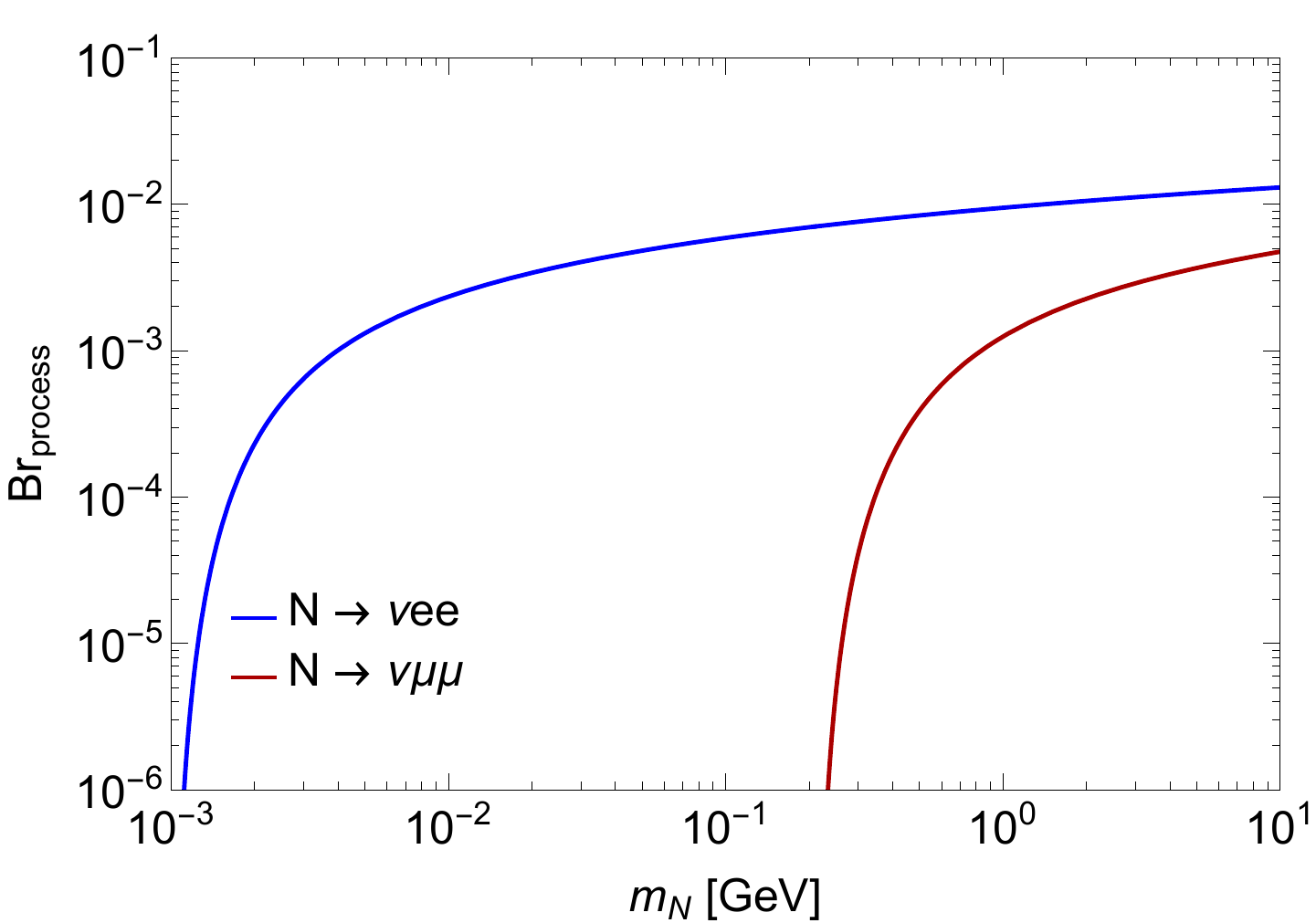}
    \caption{Branching ratios of the leptonic HNL decay processes $N \to l^{+}+l^{-}+\nu_\alpha$.}
    \label{fig:dipole-decay-br-ratios}
\end{figure}

Above the di-electron and di-muon mass threshold, the leptonic channel becomes available. This channel has been considered in ref.~\cite{Jodlowski:2020vhr} in the context of testing the dipole portal at FASER and in ref.~\cite{Arguelles:2021dqn} in the context of the T2K near detector. These processes are sub-dominant. However, they have the advantage that it is possible to reconstruct the decay vertex since we have two charged particles. In the limit $m_{N} \gg 2m_{l}$, the decay width behaves as
\begin{equation}
    \Gamma_{N\to \nu_\alpha l^{+}l^{-}} \approx \frac{\alpha_{\text{EM}}|d_\alpha|^{2}m_{N}^{3}}{12\pi^{2}}\left( \log\left[\frac{m_{N}^{2}}{m_{l}^{2}}\right]-3\right)
    \label{eq:dipole-leptonic-decay-width}
\end{equation}
The branching ratios of the leptonic decay modes are shown in Fig.~\ref{fig:dipole-decay-br-ratios}. At large masses $m_{N}\gg m_{l}$, the suppression of the leptonic decay width with a factor $\alpha_{\rm EM}/3\pi\simeq 10^{-3}$ compared to the photon channel~\eqref{eq:dipole-decay-width} gets partially compensated by the logarithm.

The two decay modes lead to different experimental signatures and imply different requirements for background rejection. Furthermore, if both decay channels can be observed, their ratio is a specific prediction of the model, serving as a smoking-gun signature.

\bigskip

The combination of the different production and decay processes leads to different signatures of the dipole portal at DUNE:
\begin{itemize}
\item[1.] Monophoton -- an event consisting of a single isolated photon appearing inside the detector. This type of events occur when the HNL is produced outside of the detector (via e.g. decays of mesons or by neutrino up-scatterings), then enters the detector, and decays through $N\to \nu+\gamma$.

\item[2.] Double-bang -- an event inside the detector consisting of two vertices: the one with recoil matter particles (electrons, nucleons, nuclei) and the one with a displaced monophoton or a pair of charged leptons~\cite{Atkinson:2021rnp, Coloma:2017ppo}. This type of signature appears if a HNL is produced inside the detector via neutrino up-scattering, and then travels a distance larger than the DUNE spatial resolution, which is of order $\Delta l_{\text{DUNE}} \simeq 1\text{ cm}$~\cite{DUNE:2020txw}.\footnote{Whether the recoil particle (and hence the HNL production point) would be detected depends on the recoil energy of the target particles. If it is below the DUNE energy detection threshold, it will be not visible. In this case, instead of the double-bang event, one would see a monophoton from the decaying HNL. In our current estimates, we assume ideal recoil energy reconstruction efficiency.}

\item[3.] Prompt photon/di-lepton -- a single event with a recoil electron/nucleon/nucleus, plus a lepton-antilepton pair or a photon. It occurs if an HNL is produced inside the detector and decays within $\Delta l_{\text{DUNE}}$ from its production point, such that its production and decay points cannot be resolved at DUNE. It is the main signature for heavy HNLs $m_{N}\gtrsim 1\text{ GeV}$.
\end{itemize}
Technical details about estimates of the number of events for various signatures and comparison of the different production channels are given in Appendix~\ref{app:dipole-portal}.

Depending on the signature, possible backgrounds include: scatterings $\nu + T \to \nu + \pi^{0}+X$~\cite{Schwetz:2020xra}, with subsequent decays $\pi^{0}\to 2\gamma$, where 2 photons cannot be resolved and the final state $X$ either was not (for the monophoton signature) or was detected (for the prompt HNL decays and double bang signature); processes $\nu + T \to n +X$ where $X$ has been detected while the neutron, being undetected, produced a displaced monophoton track at lengths comparable to its absorption length (for the double bang signature~\cite{Atkinson:2021rnp}). In the case of prompt events, a possible way to discriminate the background is to look at the energy of the recoil particles. Namely, considering the neutrino scatterings, this energy is typically much larger since the scattering mediators are heavy $W/Z$ which do not restrict the transferred momentum, while for the dipole portal it is a photon $\gamma$, which prefers $q^{2}\to 0$. A detailed study of background discrimination goes beyond the scope of this paper. Instead, we will show the reach of DUNE in the form of iso-contours corresponding to $N_{\text{events}} = 2$ events per 1 year of DUNE operation.

%%%%%%%%%%%%%%%%%%%%%%%%%%%%%%%%%%%%%%%%%%%%%%%%%%%%
\subsection{Discussion of results}

\begin{figure*}[!t]
    \centering
    \includegraphics[width=0.5\textwidth]{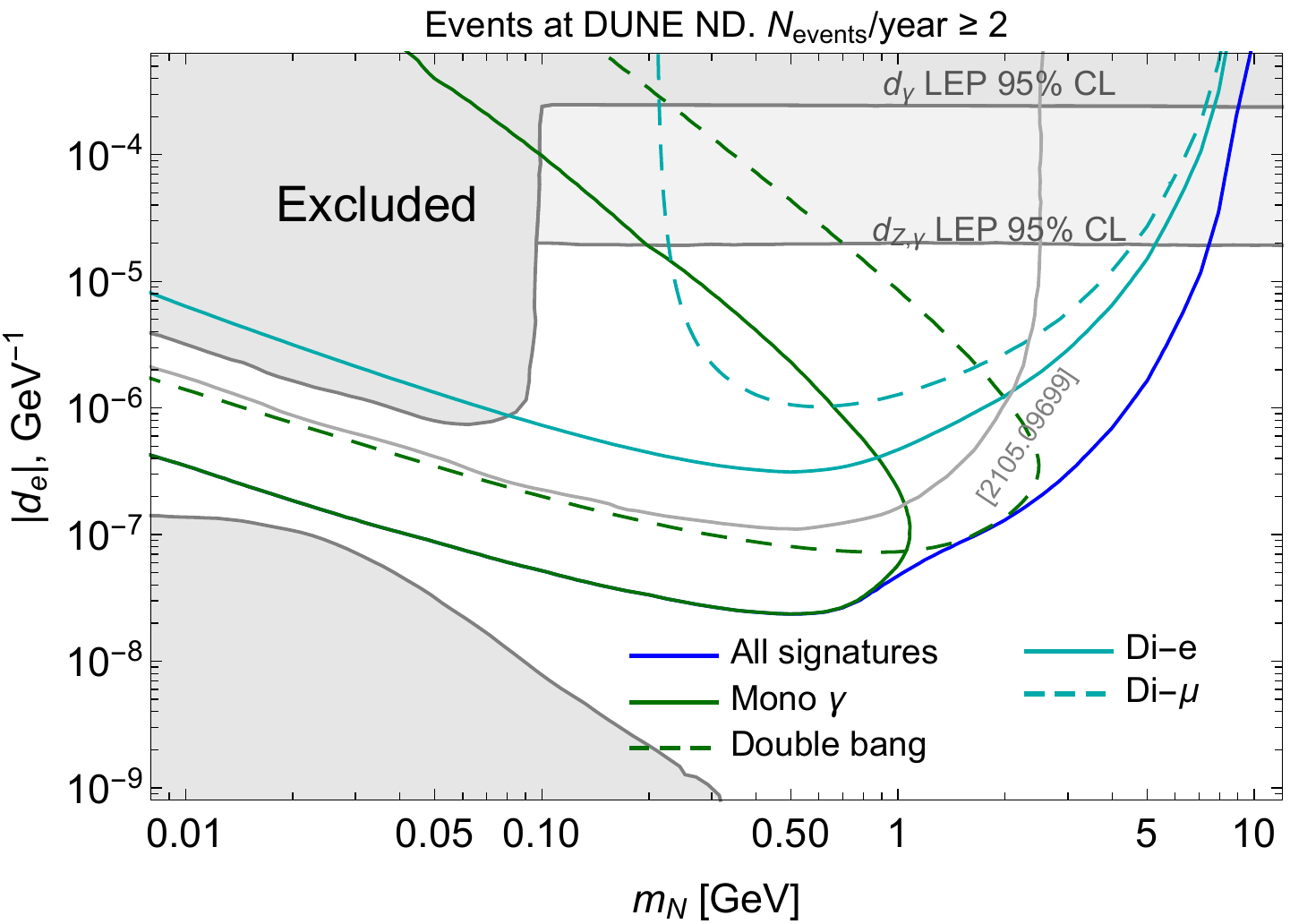}\\
    \includegraphics[width=0.5\textwidth]{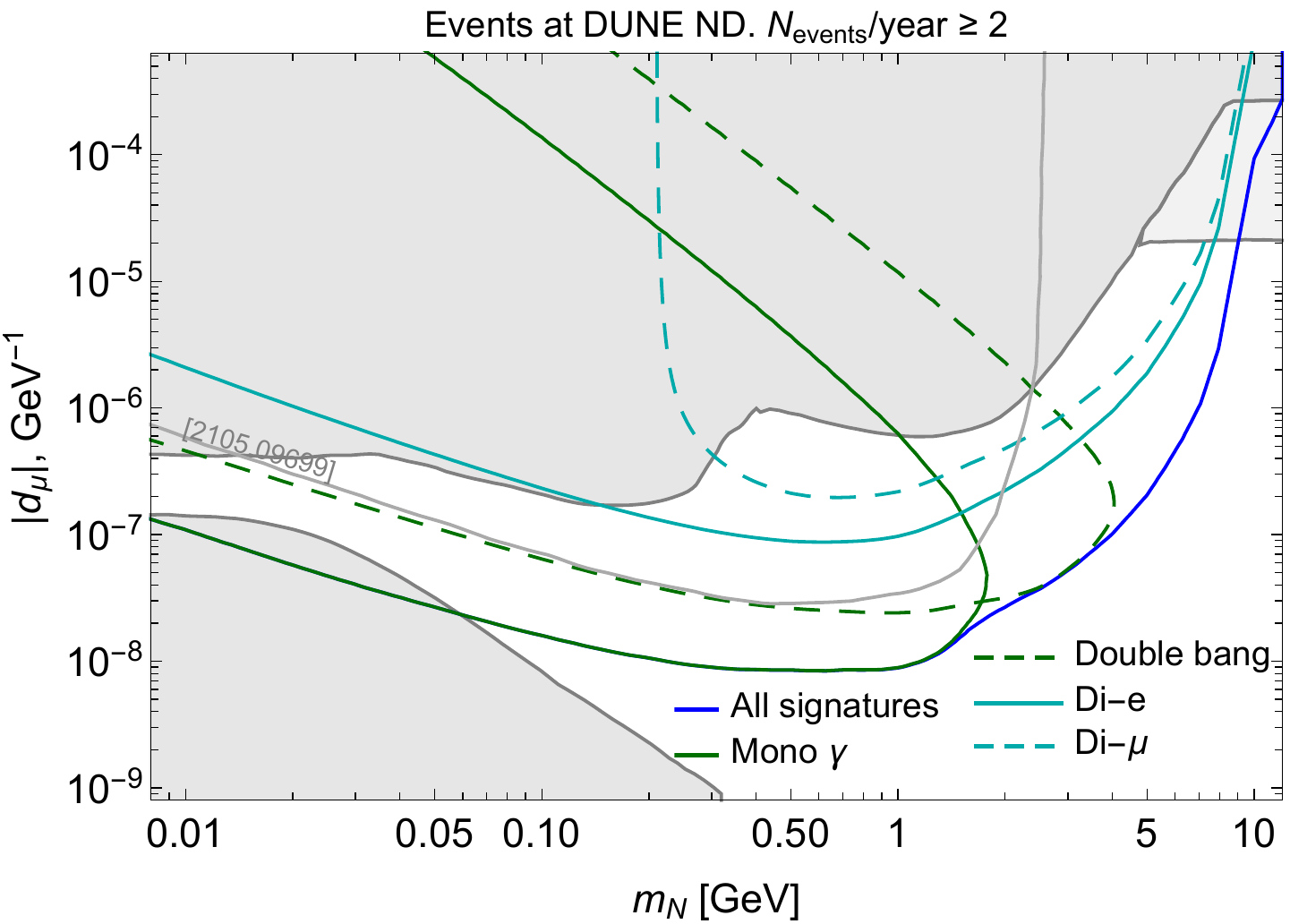}~
    \includegraphics[width=0.5\textwidth]{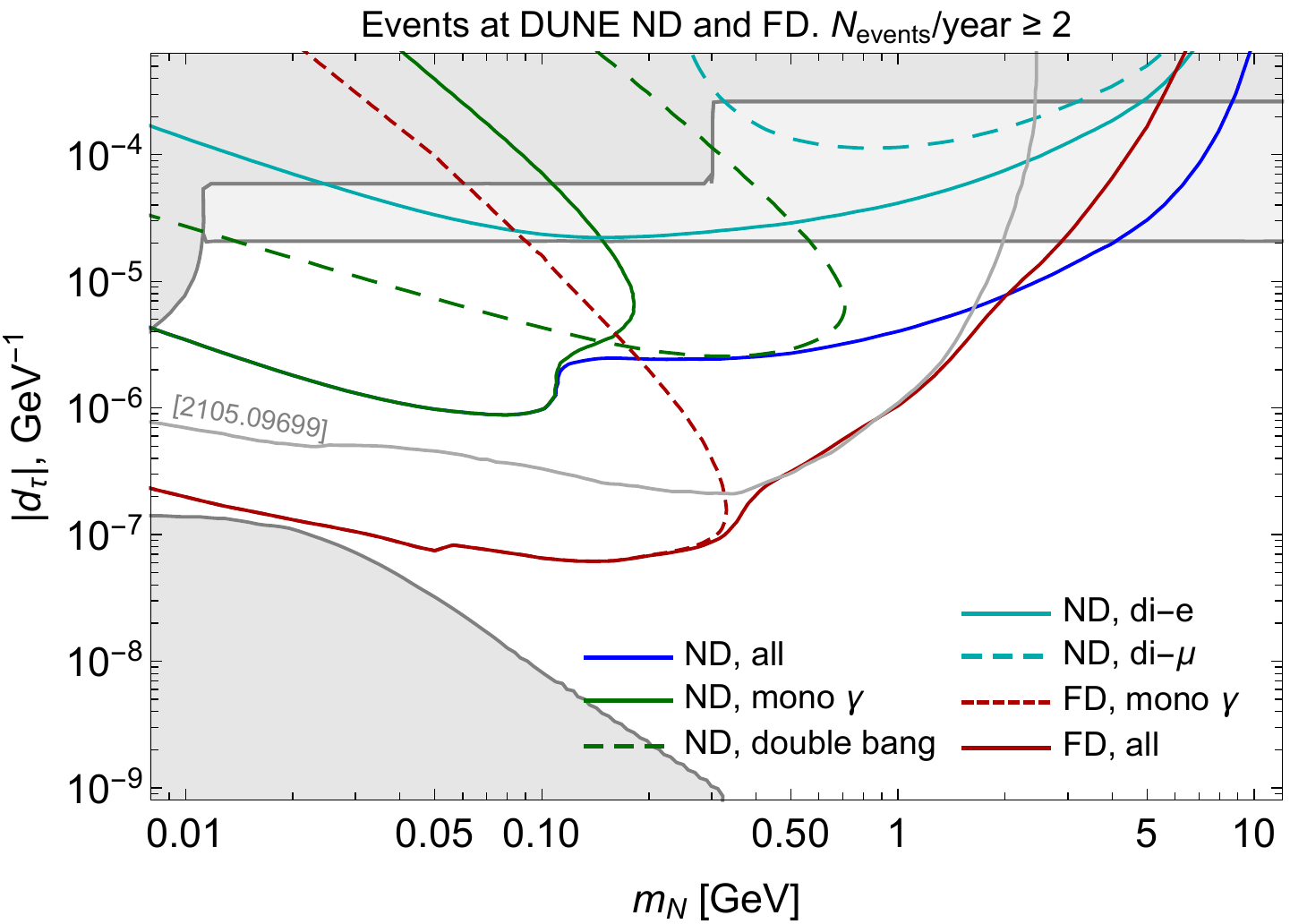}
    \caption{Sensitivity of DUNE to HNLs with dipole portal interactions with $\nu_e,\nu_\mu,\nu_\tau$ according to Eq.~\eqref{eq:dipole-portal}, from top to bottom, respectively. Iso-contours correspond to 2 events of decaying HNLs per year at the DUNE ND for $d_e$ and $d_\mu$ or ND and FD for $d_\tau$. Various signatures are considered: monophoton (green solid), ``double-bang'' (green dashed), di-leptons (red solid and red dashed), and all these signatures together (blue). The excluded parameter space is shown in gray, taken from~\cite{Schwetz:2020xra,Kamp:2022bpt}. The DUNE sensitivity obtained in~\cite{Schwetz:2020xra} corresponding to the combination of all signatures except for di-leptons is shown by the light gray curves for comparison. The bounds from LEP depend on the UV completion of the model, which we illustrate by showing the constraints assuming zero couplings to $Z$ bosons (``$d_\gamma$'') and assuming $d_{Z} = d_{\gamma}\tan(\theta_{W})$ (``$d_{Z,\gamma}$'').}
    \label{fig:dipole-portal-sensitivity}
\end{figure*}

Our sensitivity results are summarized in 
Fig.~\ref{fig:dipole-portal-sensitivity}, which shows the iso-event contours for various signatures at DUNE ND and FD assuming the CP-optimized horn configuration. (We comment on the tau-optimized configuration later in this section.)  From the figure, we conclude that depending on the signature and the flavor of the neutrino coupled to the HNL, DUNE may probe masses $m_{N}$ up to $10\text{ GeV}$. In addition, the lower bound of the sensitivity may even touch the supernova bounds. 

Let us discuss the results in more details. At the ND, the mono-photon signature (caused by decays of mesons and the outside up-scatterings) dominates at masses $m_{N}<300\text{ MeV}-1.7\text{ GeV}$, depending on the neutrino flavor coupled to $N$. The reason is that \textit{(i)} the outside medium is more dense and \textit{(ii)} more length is available for neutrinos to scatter compared to the ND (see a discussion in Appendix~\ref{app:dipole-portal-production-comparison}). For heavier HNLs, however, the number of monophoton events rapidly decreases, and the signature becomes sub-dominant. The reason for this is that the HNL lifetime shortens with increasing HNL mass, $l_{\text{decay,N}} = c\tau_{N}\gamma_{N}\propto m_{N}^{-4}d^{-2}$, and HNLs that are produced outside the detector decay before reaching the ND. To increase $l_{\text{decay,N}}$, one needs to decrease $d^{2}$. However, the decrease leads to the suppression of the number of produced HNLs. As a result, the conditions of having large enough decay length and large enough production rate become inconsistent at $m_{N}\simeq 1\text{ GeV}$. 

The di-lepton signatures, with both prompt and displaced pairs of leptons, are sub-dominant due to the small branching ratio $\text{Br}(N\to l^{+}l^{-}\nu)$. However, being combined with the events with prompt photons, they may help in discriminating signal events from background, and would provide additional information about HNL properties.

The sensitivity of the DUNE FD to $d = d_{e,\mu}$ is strongly suppressed compared to the ND. This is because of the much smaller neutrino flux of $\nu_{e,\mu}$ at the FD compared to the ND, according to Eq.~\eqref{eq:ND-to-FD-electron-muon}. Therefore, we do not show the sensitivity of the FD to these couplings. 

For $d = d_{\tau}$, there is a complementarity between the ND and FD, as illustrated in Fig.~\ref{fig:dipole-portal-sensitivity} (bottom right). As we have already discussed in Sec.~\ref{sec:neutrino-flux}, $\nu_{\tau}$s at the ND are much less numerous than at the FD, but much more energetic on average. As a result, the FD allows probing much smaller couplings than the ND for masses $m_{N}\lesssim 1\text{ GeV}$, where the energy of neutrinos is not important. However, for $m_{N}\gtrsim 1\text{ GeV}$, where the HNL production requires large neutrino energies, the sensitivity of FD quickly drops. In contrast, this is not the case for ND, which allows probing much larger masses in the domain that is currently not excluded by past experiments.

An important remark has to be made. The region of new parameter space of heavy HNLs that may be probed by DUNE is model-dependent, especially for the case of the coupling to $\nu_{\tau}$. Namely, bounds from LEP  depend on the UV completion of the model~\eqref{eq:dipole-portal}. In particular, a class of the completions introduces a coupling $d_{Z}$ to $Z$ bosons, which is related to $d$ in Eq.~\eqref{eq:dipole-portal} via the algebra of electroweak symmetry generators~\cite{Magill:2018jla}. In Fig.~\ref{fig:dipole-portal-sensitivity}, we indicate the model dependence of the bounds by showing the LEP constraints assuming $d_{Z} = 0$ and $d_{Z} \neq 0$. If $d_{Z} \neq 0$ and assuming coupling to $\nu_{\tau}$, DUNE may only probe HNLs with masses $m_{N}\lesssim 4\text{ GeV}$, while if $d_{Z} = 0$, the sensitivity extends to $m_{N}\simeq 9\text{ GeV}$.

For comparison we show in Fig.~\ref{fig:dipole-portal-sensitivity} also the sensitivity from Ref.~\cite{Schwetz:2020xra}, which combines all signatures except for the di-lepton one. We can appreciate the improved sensitivity due to additional production channels and the inclusion of the high-energy tail of the neutrino flux.\footnote{Apart from the analysis improvements discussed above, we have also corrected a few numerical issues in the calculations of Ref.~\cite{Schwetz:2020xra}, the most relevant being (i) the treatment of the off-axis neutrino flux and (ii) correcting the number of nucleon number density in the detector (roughly a factor 3). Item (i) explains different results concerning outside up-scattering production both for ND and FD sensitivities (see a brief discussion in Appendix~\ref{app:dipole-portal-production-comparison}).
%The different results are related to the inclusion of the neutrino flux files in the GLoBES format~\cite{DUNEfluxes} in Ref.~\cite{Schwetz:2020xra}, whereas in our current analysis we depart directly from the DUNE simulation files~\cite{DUNEfluxes}, see Appendix~\ref{app:neutrino-meson-flux} for details.
} 
We note also that when adopting the same assumptions about exposure and experimental parameters, our results for the double-bang signature are in good agreement with Ref.~\cite{Atkinson:2021rnp}.

Above, we have considered the HNLs as the Dirac particles. The analysis for the Majorana case is completely similar. The only difference in the sensitivity is purely numerical: given the same coupling $d$, the Majorana HNLs would have twice larger decay width $\Gamma_{N}$. The effect on the sensitivity depends on the considered parameter space and the signature. Namely, in the regime where the HNL decay length is much larger than the scale of DUNE, the number of events is $N_{\text{events}}\propto d^{2}\Gamma_{N}$, hence being twice larger for Majorana HNLs. Since $\Gamma_{N}\propto d^{2}$, in the plane $d-m_{N}$, the lower bound of the sensitivity would be $2^{1/4}$ lower for the Majorana particles. In the opposite regime of short decay lengths, the number of events behaves as $N_{\text{events}}\propto d^{2}\times \exp[-l_{\text{min}}\Gamma_{N}/c\gamma_{N}]$, where $l_{\text{min}}$ is the minimal displacement for the given signature. For the prompt signatures, $l_{\text{min}} = 0$, and the decay width does not affect the sensitivity at all (the case of prompt signatures). This is not the case for the case of the outside production and the double bang. The upper bound of the sensitivity to these signatures would be a factor of $\sqrt{2}$ lower for Majorana particles. Finally we note that apart from these numerical differences, the presence of a Majorana mass for the HNL would lead to theoretical inconsistencies, as in this case, the dipole operator with $d$ in the relevant range would induce a large contribution to the light neutrino mass matrix via a loop diagram, in conflict with bounds on neutrino masses, see e.g.,~\cite{Magill:2018jla}. Therefore, a consistent theory of large $d$ in case of Majorana NHLs would require more elaborate model building.

\bigskip

\begin{figure*}[!t]
    \centering
    \includegraphics[width=0.45\textwidth]{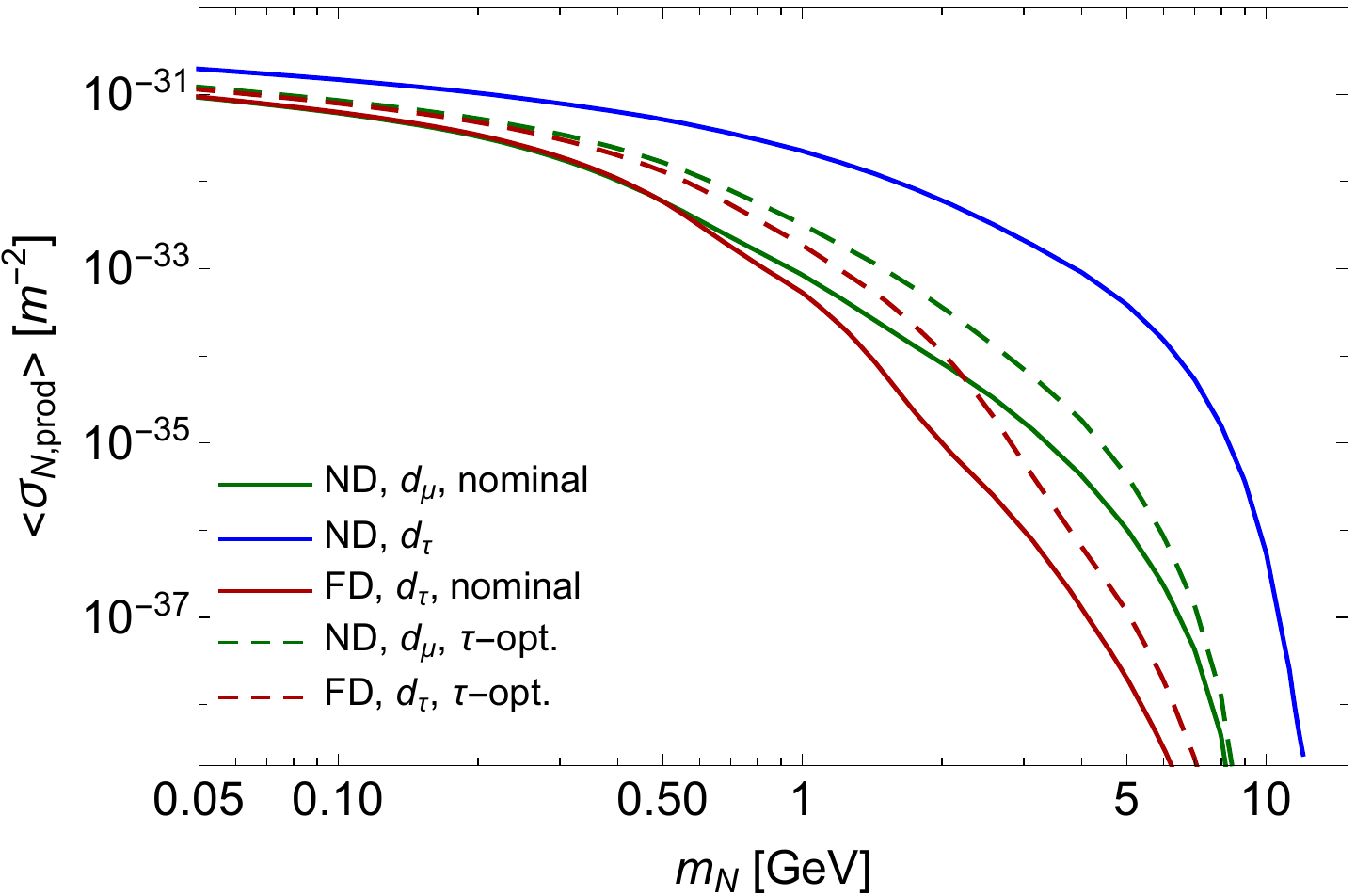}~\includegraphics[width=0.45\textwidth]{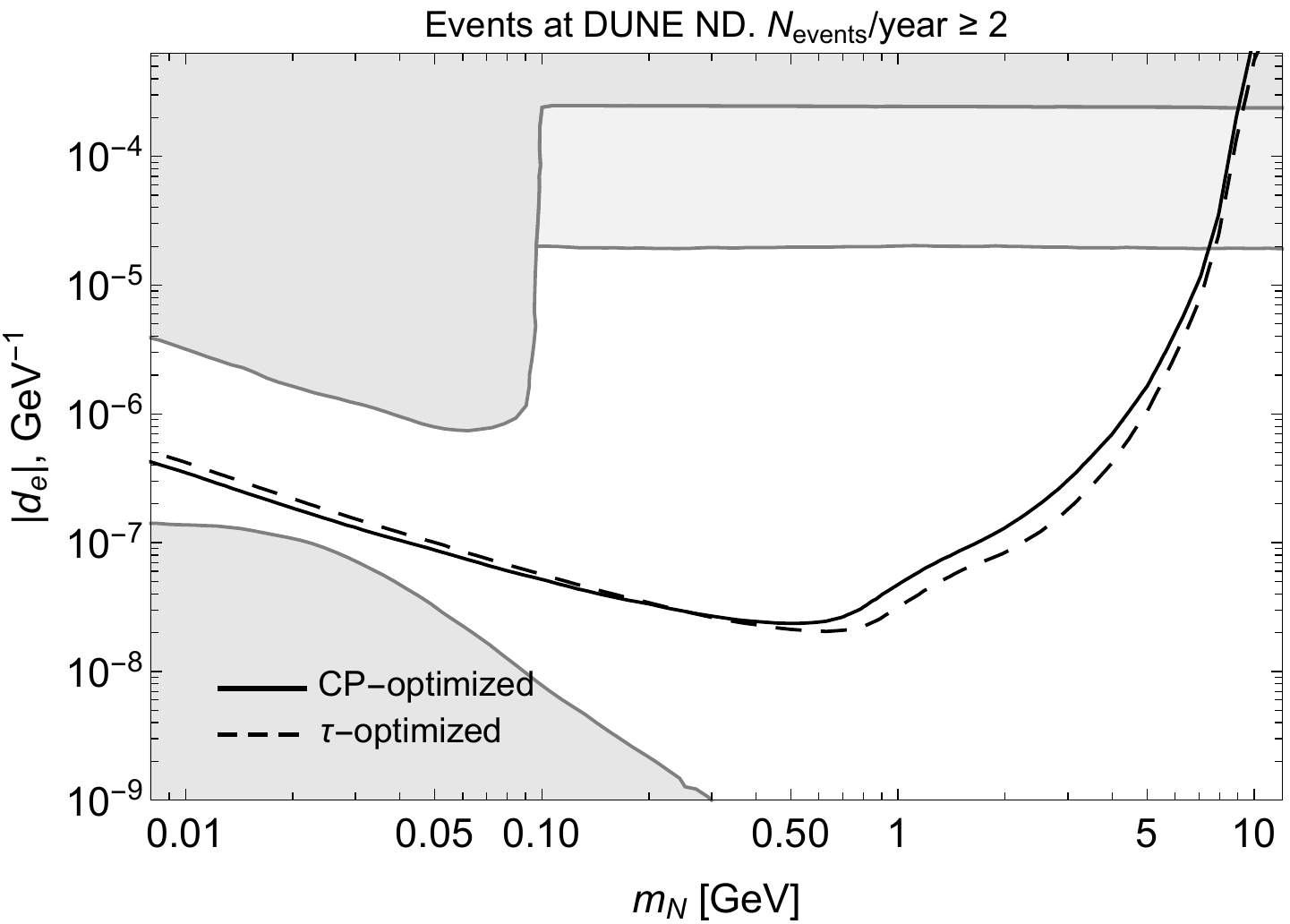}\\ \includegraphics[width=0.45\textwidth]{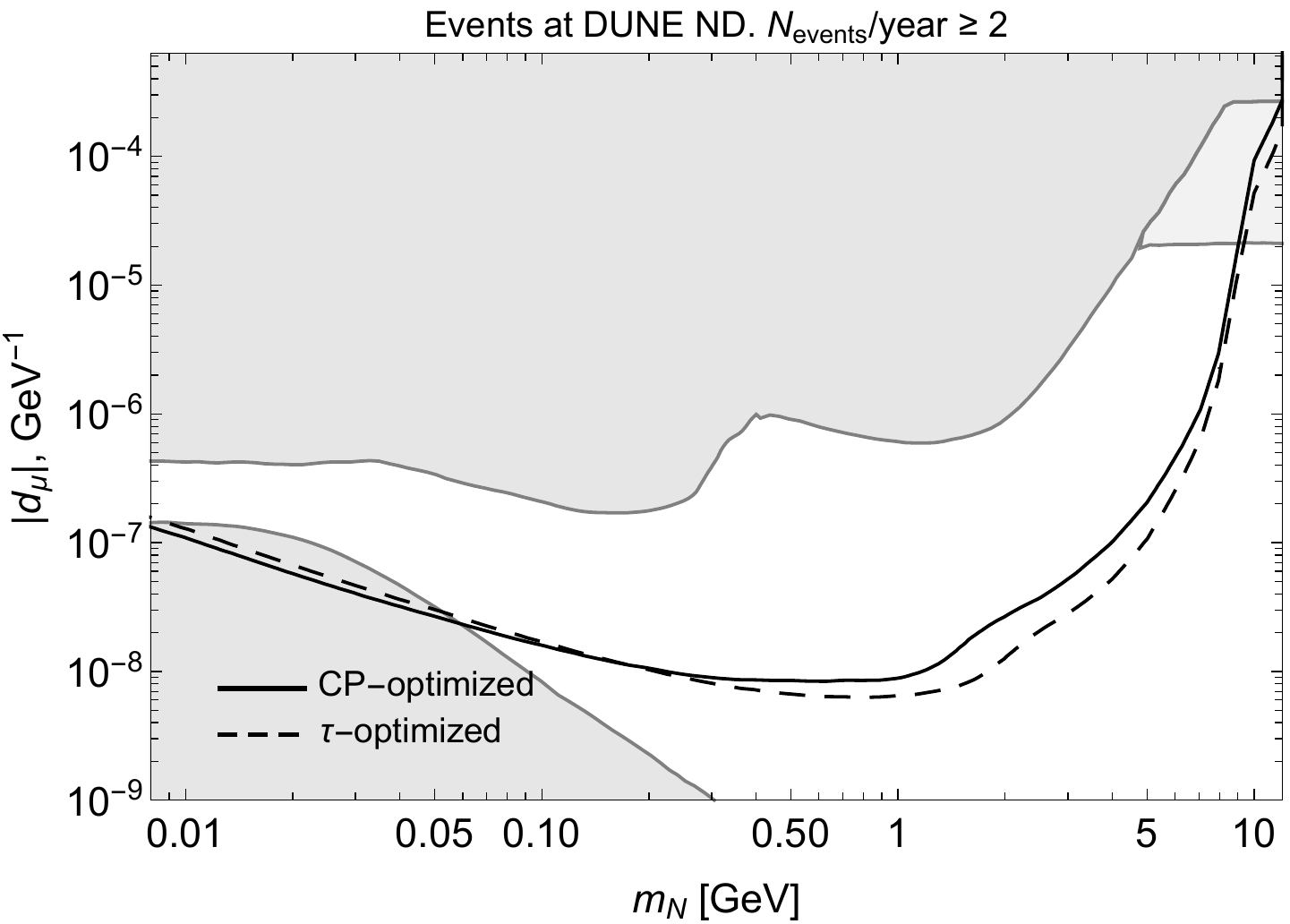}~\includegraphics[width=0.45\textwidth]{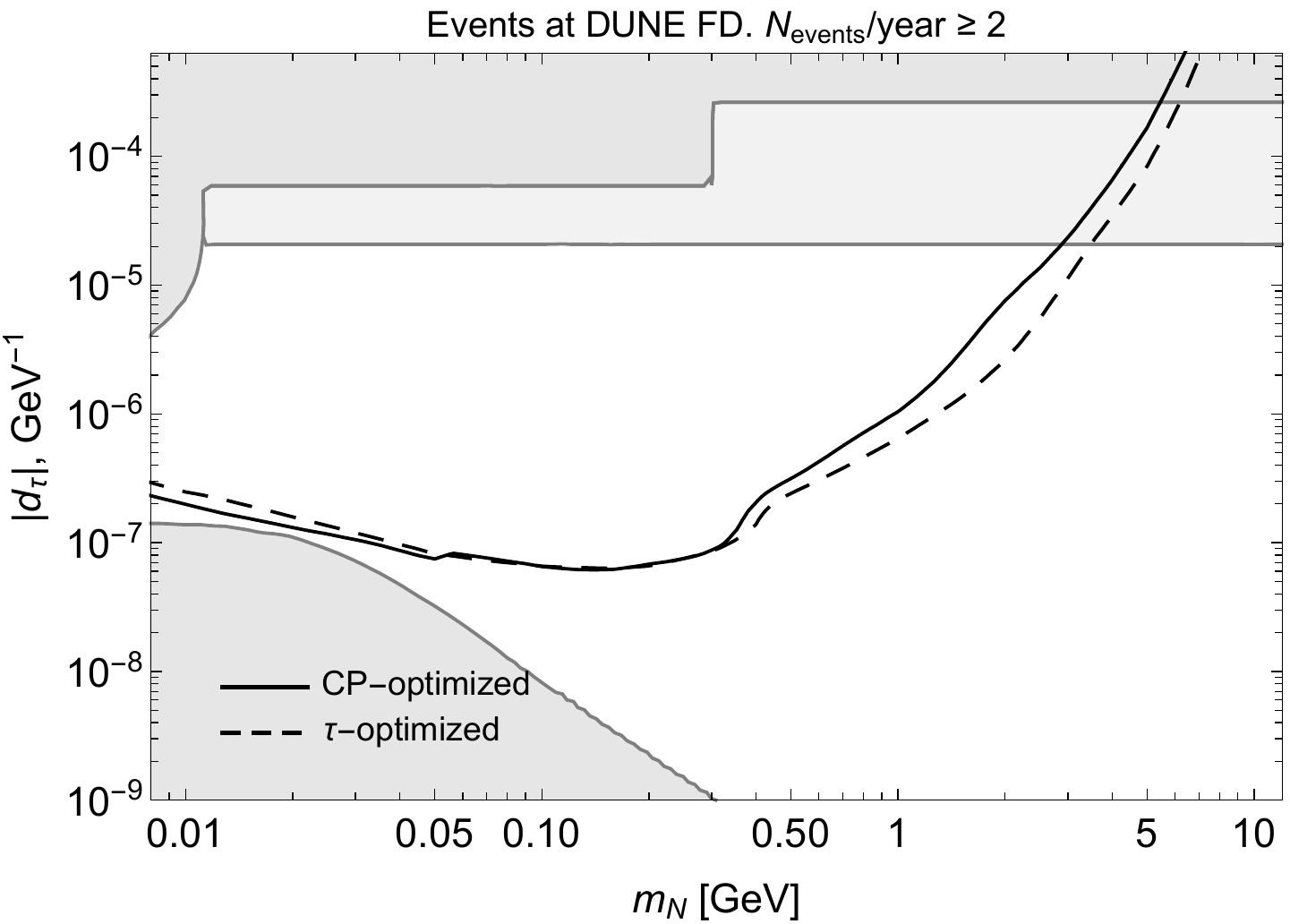}
    \caption{Sensitivity of DUNE to the dipole portal considering two different horn configuration modes: CP-optimized and tau-optimized. \textit{Top left panel}: total HNL production cross sections averaged over neutrino energies (Eq.~\ref{eq:dipole-cross-sections-averaged}) for $d = d_{\mu} = 1\text{ GeV}^{-1}$ and $d = d_{\tau} = 1\text{ GeV}^{-1}$, assuming different horn configurations. \textit{Remaining panels}: comparison of the 2-event per year iso-contour curves for the two horn configurations for $d_e$ (top right), $d_\mu$ (bottom left) and $d_\tau$ (bottom right). }
    \label{fig:DUNE-CP-optimized-tau-optimized}
\end{figure*}

Let us now consider the impact of the different horn configurations, in particular the tau-optimized flux. The fluxes of all the neutrinos except for $\nu_{\tau}$ at the ND are very sensitive to the horn configuration.\footnote{$\tau$ neutrinos at the ND are produced mainly by decays of $D_{s}$ mesons and $\tau$ leptons, which occur almost instantly after the production of these particles inside the target. Therefore, the $\nu_{\tau}$ flux is insensitive to the horn configuration.} In Fig.~\ref{fig:neutrino-fluxes}, we compare the fluxes for the CP-optimized and tau-optimized configurations. We see that the amount of neutrinos with energies $E_{\nu}>5\text{ GeV}$ for the tau-optimized flux may be larger by a factor up to $\simeq 20$. This translates to a comparable increase of the cross section of the production of heavy HNLs with $m_{N}\gtrsim 2\text{ GeV}$, which requires high-energy neutrinos, see Fig.~\ref{fig:DUNE-CP-optimized-tau-optimized} (top left).
The comparison of the sensitivities of DUNE assuming the CP-optimized and tau-optimized horn configurations is shown in the remaining panels of Fig.~\ref{fig:DUNE-CP-optimized-tau-optimized}. In dependence on the HNL mass, the improvement in the sensitivity may reach a factor as large as 3.

%%%%%%%%%%%%%%%%%%%%%%%%%%%%%%%%%%%%%%%%%%%%%%%%%%%%%%%
\section{Neutrinophilic scalar portal}
\label{sec:neutrinophilic-portal}
%%%%%%%%%%%%%%%%%%%%%%%%%%%%%%%%%%%%%%%%%%%%%%%%%%%%%%%

The neutrinophilic scalar portal is characterized by the effective Lagrangian~\cite{Berryman:2018ogk, Kelly:2019wow}
\begin{equation}
   \mathcal{L}_{\text{neutrinophilic}} = \frac{g_{\alpha}}{2}\phi \overline{\nu_{\alpha}^{c}}\nu_{\alpha} +\text{H.c.},
   \label{eq:neutrinophilic-portal}
\end{equation}
where $\nu^{c}$ means charged conjugated neutrino field, and $\phi$ is a massive complex scalar carrying lepton number $-2$. The interaction in Eq.~\eqref{eq:neutrinophilic-portal} can emerge from a dimension-6 operator of the form $(LH)(LH)\phi/\Lambda^2$, with $L$ and $H$ denoting the lepton and SM Higgs doublet fields, respectively. Again we remain agnostic about the UV completion and focus on the low-energy implications of the effective interaction \eqref{eq:neutrinophilic-portal}. The scalar introduced in Eq.~\eqref{eq:neutrinophilic-portal} can also act as a portal to dark matter \cite{Kelly:2019wow}. SM precision measurements of meson decays or invisible Higgs decays provide constraints on the parameters  $m_{\phi}$ and $g_{\alpha}$ \cite{Berryman:2018ogk}; the excluded range is shown by the gray shaded region in Fig.~\ref{fig:neutrinophilic-portal-sensitivity}. 

\begin{figure}[t]
    \centering
    \includegraphics[width=0.45\textwidth]{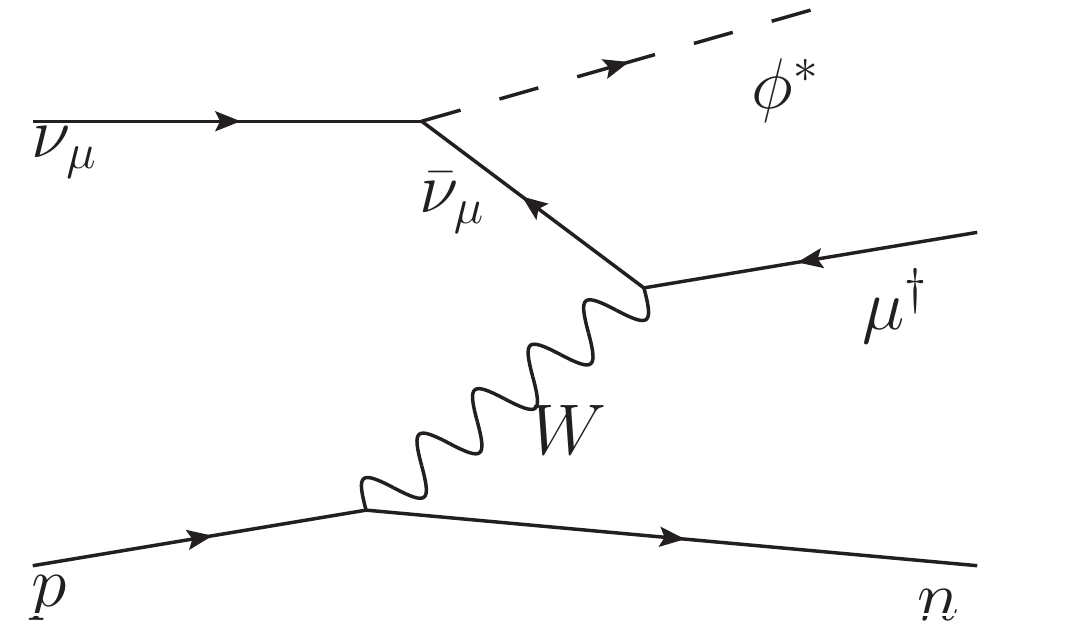}
    \caption{The production diagram of a scalar particle $\phi$ via the neutrinophilic portal, Eq.~\eqref{eq:neutrinophilic-portal}.}
    \label{fig:neutrinophilic-production}
\end{figure}

At neutrino experiments, the neutrinophilic scalar portal in Eq.~\eqref{eq:neutrinophilic-portal} may be searched for by an excess of events with missing transverse momentum $p_{T}$~\cite{Kelly:2019wow,Coyle:2022bwa}, e.g.
\begin{equation}
    \nu_{\mu} + p^{+} \to \phi^{\star} + \mu^{+} + X,
    \label{eq:neutrinophilic-signature}
\end{equation}
where $X$ is a hadronic final state, and $\phi$ leaves the detector invisibly or decays into neutrinos and hence leads to missing $p_{T}$ (see Fig.~\ref{fig:neutrinophilic-production}).
Indeed, since neutrinos are collimated with respect to the beam axis, they carry negligible $p_{T}$. Therefore, the products of the reaction~\eqref{eq:neutrinophilic-signature} have total zero $p_{T}$. Finite reconstruction efficiencies smear the $p_{T}$ distribution even if all the reaction products are detected. However, the resulting distribution is softer than the distribution of events with invisible particles such as $\phi$. Additionally, due to the lepton number violation in the operator~\eqref{eq:neutrinophilic-portal}, an antimuon is produced together with hadrons. If the muon charge can be identified, this would allow reducing background even further. In particular, the study performed in~\cite{Kelly:2019wow} has shown that the distribution of $p_{T}$ of muons and nucleons produced by the process $\nu + n \to p^{+} + \mu^{-}$, assuming the muon charge identification, drops to zero at $p_{T}\simeq 0.5\text{ GeV}$. 

In~\cite{Kelly:2019wow}, the sensitivity of DUNE to the neutrinophilic scalar portal has been estimated by considering the quasi-elastic (QE) scattering
\begin{equation}
    \nu_{\mu} + p^{+} \to n + \phi + \mu^{+},
    \label{eq:neutrinophilic-quasi-elastic}
\end{equation}
requiring that the missing transverse momentum is $p_{T,\text{miss}}>0.5\text{ GeV}$. 
In this work, we extend the analysis of \cite{Kelly:2019wow} in the following ways. First, we include the high-energy tail of the neutrino flux, by considering also neutrinos with  $E_{\nu} > 10\text{ GeV}$. This will provide sensitivity to somewhat larger scalar masses 
(remind Fig.~\ref{fig:Enumin}). Second, we include form-factors to the QE proton or neutron vertex~\cite{Leitner:2009zz} in order to take into account that especially for heavy $\phi$, $m_{\phi}\gtrsim r_{p}^{-1} \simeq 1\text{ GeV}$, the nucleons cannot be considered as point-like, leading to a suppression of the QE cross section. Third, we consider in addition the production process via DIS, which dominates when the incoming neutrinos are energetic enough.
In Appendix~\ref{app:neutrinophilic-portal} we provide details on the relevant cross sections and further discussion of the respective impact of each of these improvements in the analysis.

The 90\% CL sensitivity to the coupling $g_{\mu}$, considering the scalar interactions with $\nu_{\mu}$, is shown in Fig.~\ref{fig:neutrinophilic-portal-sensitivity}. We require $p_{T,\phi} > 0.5\text{ GeV}$ and assume the absence of background, which, according to Ref.~\cite{Kelly:2019wow}, corresponds to the case when the muon charge may be identified (see also the discussion above). Therefore, the sensitivity is calculated by requiring $N_{\text{events}}>2.3$ for the full DUNE exposure of 5 years in the neutrino mode ($5.5\times10^{21}$ PoT). If assuming no muon charge ID, the 90\% sensitivity would drop by a factor of $\simeq 2$ because of the non-zero background~\cite{Kelly:2019wow}.

In this figure, we also show the ND sensitivity from Fig.~2 in~\cite{Kelly:2019wow}, where we choose the curve obtained under the assumption of the presence of the muon charge identification. As for the cases of the interactions with $\nu_e$ and $\nu_\tau$, DUNE has no sensitivity. There are two reasons for this. First, in case of interactions with $\nu_{e,\tau}$, electrons and $\tau$ leptons would be produced in the process~\eqref{eq:neutrinophilic-quasi-elastic} instead of muons. It is more difficult to look for missing $p_{T}$ in this case, since it is more complicated to reconstruct the kinematics of electrons and $\tau$ (which in addition decays into $\tau$ neutrinos). Second, even if assuming a perfect kinematics reconstruction of $e$ and $\tau$, the fluxes of $\nu_{e,\tau}$ are suppressed by at least two orders of magnitude (see Fig.~\ref{fig:neutrino-fluxes}). This translates into at least one order of magnitude suppression of the sensitivity.  On the other hand, the currently excluded domain for the $e,\tau$ cases remains very similar to the case of the interaction with $\nu_{\mu}$ or even becomes tighter (see~\cite{Kelly:2019wow}).

\begin{figure}[t]
    \centering
    \includegraphics[width=0.45\textwidth]{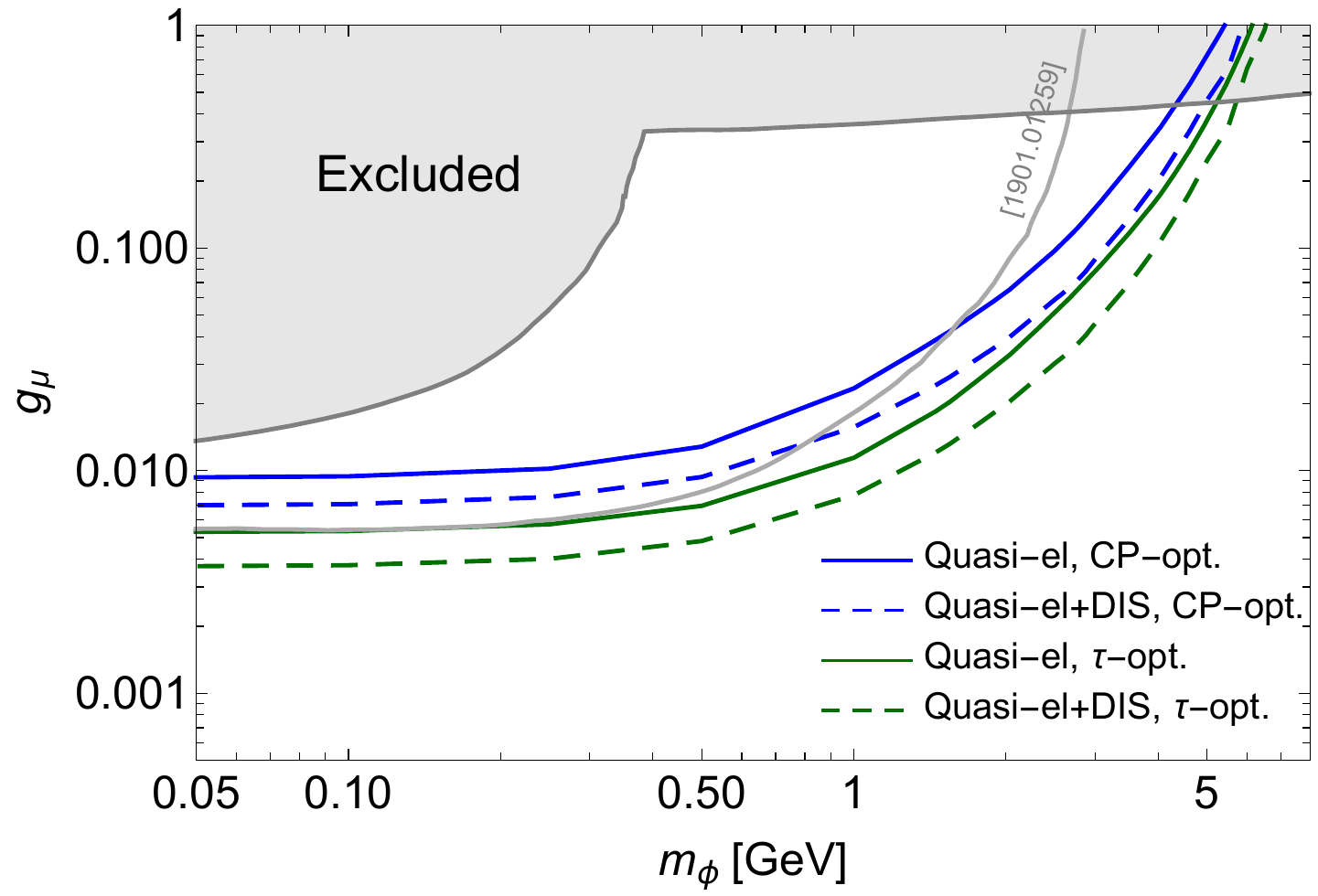}
    \caption{90\% C.L. sensitivity ($N_{\text{events}}>2.3$ in background-free regime) of the DUNE ND to the a neutrinophilic scalar interacting with $\nu_\mu$, see Eq.~\eqref{eq:neutrinophilic-portal}. Two horn configurations are considered: CP-optimized (blue curves), and tau-optimized (green curves). Solid curves show the sensitivity assuming the production of the neutrinophilic scalar $\phi$ by only the quasi-elastic process, Eq.~\eqref{eq:neutrinophilic-quasi-elastic}, while the dashed colored lines correspond to the production by QE and DIS scattering combined. For comparison, the gray line shows the sensitivity obtained in Ref.~\cite{Kelly:2019wow} for the CP-optimized horn configuration.}
    \label{fig:neutrinophilic-portal-sensitivity}
\end{figure}

Let us briefly summarize the results. Compared to the estimate of the sensitivity from~\cite{Kelly:2019wow}, which has been performed for the CP-optimized horn configuration, at small masses $m_{\phi}\lesssim 1.5-2\text{ GeV}$, our estimate shows slightly worse sensitivity. This is the effect of the inclusion of the nucleon form factors. The inclusion of DIS can only partially compensate for this effect, see blue solid and blue dashed curves.
However, due to the inclusion of high-energy neutrinos, we find that DUNE has a sensitivity to somewhat larger scalar masses, up to $m_{\phi} \simeq 4.5\text{ GeV}$.
Compared to the CP-optimized horn configuration, the tau-optimized configuration improves the sensitivity for the whole mass range (green curves). In addition, it extends the probed mass region by $\simeq 1\text{ GeV}$.

The point-like nucleon approximation may also affect the analysis of background performed in~\cite{Kelly:2019wow}, according to which the SM background vanishes at $p_{T}>0.5\text{ GeV}$. In particular, form factors suppress large transverse momentum of nucleons (and hence muons), which means that in reality the SM background is even softer and drops to zero at smaller $p_{T}$. Therefore, the cut $p_{T,\text{miss}}>0.5\text{ GeV}$ for the signature~\eqref{eq:neutrinophilic-signature} may be relaxed. This question is a subject of a separate study; however, we have checked that the relaxation of the cut to $p_{T,\text{miss}}>0.2\text{ GeV}$ may improve the sensitivity by up to a factor of 2.

%%%%%%%%%%%%%%%%%%%%%%%%%%%%%%%%%%%%%%%%%%%%%%%%%%%%%%%%%%%%%%%%
\section{Conclusions}
%%%%%%%%%%%%%%%%%%%%%%%%%%%%%%%%%%%%%%%%%%%%%%%%%%%%%%%%%%%%%%%%
\label{sec:conclusions}

In this paper, we have estimated the sensitivity of the DUNE experiment to new physics particles that interact with neutrinos, considering the neutrino dipole and neutrinophilic scalar portals as examples. Compared to previous studies, we have included new production channels, considered various detection signatures, and taken into account the high-energy tail of the DUNE neutrino flux.  
Although the fraction of neutrinos with higher energies is strongly suppressed, neutrinos with  $E_{\nu}\gtrsim 10\text{ GeV}$  are important for the production of heavy particles with masses $m\gtrsim 1\text{ GeV}$, see Sec.~\ref{sec:neutrino-thresholds}.
We have demonstrated that the inclusion of high-energy neutrinos may significantly improve the sensitivity, extending the probed mass range. In particular, DUNE may probe the dipole portal up to HNL masses $m\simeq 9\text{ GeV}$ (Fig.~\ref{fig:dipole-portal-sensitivity}), and the neutrinophilic scalar portal up to masses $m\simeq 5.5\text{ GeV}$ (Fig.~\ref{fig:neutrinophilic-portal-sensitivity}), with some dependence on the horn configuration.

For the dipole portal, we have studied in detail various production mechanisms and signatures in the detector. Depending on the parameter region, the mono-photon signal, double-bang signature, or a single event with a shower + prompt photon or lepton-pair may be observable. The relative size of these signals will be a smoking gun signature to identify the dipole portal model. Furthermore, different signatures may require dedicated analysis cuts and background mitigation strategies.

\section*{Acknowledgments}
We thank Nilay Bostan and Laura Fields for useful discussions on the neutrino flux at DUNE and for providing the simulation datasets with mesons and neutrinos at DUNE, and Kevin Kelly for discussions on the sensitivity of DUNE to the neutrinophilic portal. We are also grateful to Albert Zhou for his involvement in the early stage of this work. Jing-yu Zhu is supported in part by the China and Germany Postdoctoral Exchange Program from the Office of China Postdoctoral Council and the Helmholtz Centre under Grant No. 2020031 and by the National Natural Science Foundation of China under Grant No. 11835005 and 11947227. This project has received support from the European Union’s Horizon 2020 research and innovation program under the Marie Sklodowska-Curie grant agreement No. 860881-HIDDeN.

\bibliography{bib.bib}

\newpage

\appendix

\onecolumngrid

\section{Flux of muon and electron neutrinos at DUNE}
\label{app:neutrino-meson-flux}

\begin{figure}[t!]
    \centering
    \includegraphics[width=0.5\textwidth]{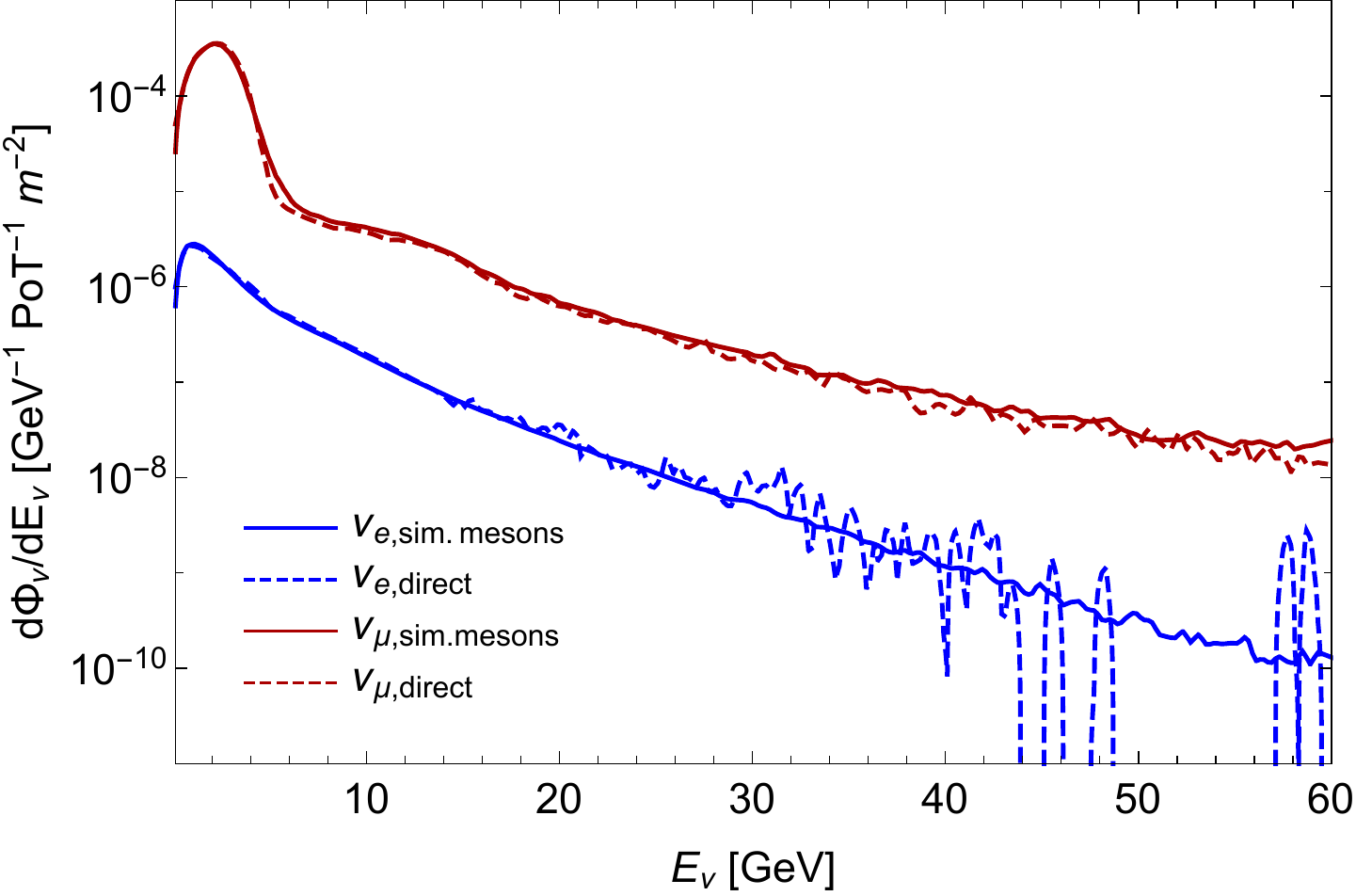}
    \caption{Electron and muon neutrino fluxes at DUNE ND. The solid lines show the fluxes obtained by regenerating meson decays into $\nu_{e,\mu}$, while the dashed lines show the fluxes extracted directly from the DUNE simulation files (see text for details).}
    \label{fig:neutrino-flux-cross-check}
\end{figure}

In order to get the flux of $\nu_{e}$ and $\nu_{\mu}$ neutrinos and the distribution of decaying mother mesons, we have used the publicly available results of the detailed GEANT4~\cite{GEANT4:2002zbu, Allison:2016lfl, Allison:2006ve} based simulation (G4LBNF) of the LBNF beamline developed by the DUNE collaboration~\cite{DUNE:2020lwj, DUNEfluxes}. It is possible to extract the flux of the electron and muon neutrinos from the simulations at the DUNE ND and FD directly from the simulation files, see Fig.~\ref{fig:neutrino-flux-cross-check}. In the simulation data, there are more than $12$ million of decaying pions and 1 million of decaying kaons. These numbers are enough to produce the high-energy spectrum for $\nu_{\mu}$ very well. However, the number of simulated decays into electron neutrinos is a few orders of magnitude smaller, and this is especially the case for the high-energy tail. There are two reasons for this. First, most of the muons ($\simeq 99\%$) from $\pi$, instead of decaying, get absorbed outside the decay pipe because of their long lifetime. The decaying muons typically have small energies $E_{\mu}\lesssim 10\text{ GeV}$, which means that they mostly produce low-energy neutrinos. Second, in the DUNE simulation, only a fraction of kaons, namely, $\text{Br}(K \to \nu_{e})/\text{Br}(K \to \nu_{\mu})$, has been used to simulate decays into electron neutrinos. This results in large wiggles in the obtained distribution for energies $E_{\nu}\gtrsim 30\text{ GeV}$.
 
Instead of using this direct flux, we have re-simulated the decays of $\pi, K,\mu$ into $\nu_{e}$, allowing all the kaons can decay into neutrinos, and then selecting the neutrinos that point to the ND. To simulate the decay, we first extracted the momenta and decay coordinates of the mesons and muons from the DUNE simulation dataset, then generated the phase space of their decay products at the rest frame of the decaying particles, and then boosted to the lab frame. 

We have used the following matrix elements:
\begin{align}
    \mathcal{M}_{\mu^{+} \to \bar{\nu}_{\mu}+\nu_{e}+e^{+}} \approx & \frac{G_{F}}{\sqrt{2}}\bar{v}(p_{\mu})\gamma_{\mu}(1-\gamma_{5})v(p_{\nu_{\mu}}) \times \bar{u}(p_{\nu_{e}})\gamma^{\mu}(1-\gamma_{5})v(p_{e}), \\
    \mathcal{M}_{\pi^{+}/K^{+} \to l^{+} +\nu_{l}} \approx & \frac{G_{F}f_{\pi/K}}{\sqrt{2}} p_{\pi/K,\mu} \times \bar{u}(p_{\nu_{l}})\gamma^{\mu}(1-\gamma_{5})v(p_{l}), \\
    \mathcal{M}_{K^{+/0} \to l^{+} +\nu_{l}+\pi^{0/-}} \approx & \frac{G_{F}V_{us}}{\sqrt{2}} \bar{u}(p_{\nu_{l}})\gamma^{\mu}(1-\gamma_{5})v(p_{l}) \times 
    [(p_{\pi}+p_{K})_{\mu}f_{+}
    (p_{K}-p_{\pi})^{2}+(p_{\pi}-p_{K})_{\mu}f_{-}(p_{K}-p_{\pi})^{2}]
    \end{align}
for the muons, pions, and kaons~\cite{Bijnens:1994me} correspondingly, where $f_{\pm}$ are $K\to \pi$ transition form-factors~\cite{Aoki:2017spo}. As a cross-check, we have verified that the phase space distribution of particles in the process $\mu^{+} \to e^{+}+\nu_{e}+\bar{\nu}_{\mu}$ at the rest frame of decaying $\mu^{+}$ generated by us coincides with that generated by MadGraph5~\cite{Alwall:2014hca}.

The comparison of the obtained fluxes with the directly extracted fluxes is shown in Fig.~\ref{fig:neutrino-flux-cross-check}. They agree with each other very well at all energies in consideration, while the regenerated flux for $\nu_{e}$ has fewer wiggles at high energies. Small differences at energies $E_{\nu}\lesssim 10\text{ GeV}$ may be explained by slightly different sizes and geometry of the ND considered in the DUNE simulations and in our estimates. 

\section{Dipole portal}
\label{app:dipole-portal}

\subsection{Production}
\label{app:dipole-portal-production}
\subsubsection{Decays of mesons}
\label{app:dipole-portal-production-mesons}

HNLs may be produced by decays of mesons via a virtual photon or neutrino. Examples are
\begin{equation}
    \pi^{0}/\eta\to \gamma+N+\nu, \quad \rho^{0}\to N+\nu,
    \label{eq:dipole-production-mesons-1}
\end{equation}
going through a virtual photon, and
\begin{equation}
    \pi^{\pm}/K^{\pm} \to N+\gamma+l^{\pm},
    \label{eq:dipole-production-mesons-2}
\end{equation}
going through a virtual neutrino,
see Fig.~\ref{fig:dipole-portal-production-diagrams}.

The Lagrangians describing the SM vertices of the decays~\eqref{eq:dipole-production-mesons-1},~\eqref{eq:dipole-production-mesons-2} are:
\begin{equation}
\mathcal{L}_{m^{0}\gamma\gamma} = g_{m^{0}\gamma\gamma}m^{0}F_{\mu\nu}\tilde{F}^{\mu\nu}, \quad \tilde{F}_{\mu\nu} = \frac{1}{2}\epsilon_{\mu\nu\alpha\beta}F^{\alpha\beta}
\end{equation}
for the transition $\pi^{0}\to \gamma \gamma^{*}(\to N+\nu)$;
\begin{equation}
\mathcal{L}_{m^{-}l\nu} = f_{m^{-}l\nu}\partial_{\mu}m^{-} \bar{l}\gamma^{\mu}(1-\gamma_{5})\nu_{l}, \quad m^{-} = \pi^{-}/K^{-}
\end{equation}
for the transition $m^{-}\to l^{-}\bar{\nu}_{l}^{*}(\to N\gamma)$; and~\cite{Fujiwara:1984mp}
\begin{equation}
    \mathcal{L}_{\rho^{0}\gamma} = g_{\rho^{0}\gamma}\rho^{0}_{\mu}A^{\mu}
\end{equation}
for the transition $\rho^{0}\to \gamma^{*}(\to N\nu)$.

The values of the effective couplings in these Lagrangians are: 
\begin{equation}
g_{m^{0}\gamma\gamma} = \frac{\alpha_{\text{EM}}}{4\pi f_{m^{0}}}\approx 6.2\cdot 10^{-3}\text{ GeV}^{-1},
\end{equation}
where $f_{\pi^{0}} \approx 93\text{ MeV}$ and $f_{\eta} \approx 116\text{ MeV}$;
\begin{equation}
    f_{m^{-}l\nu} = f_{m^{-}}G_{F}V_{\text{CKM}}^{(m^{-})} \approx \begin{cases} 1.06\cdot 10^{-6}\text{ GeV}^{-1}, \quad m^{-} = \pi^{-} \\ 2.9\cdot 10^{-7}\text{ GeV}^{-1}, \quad m^{-}=K^{-},\end{cases} 
\end{equation}
with $f_{m^{-}}$ being the decay constant ($f_{\pi^{-}} \approx 93\text{ MeV}$, $f_{K^{-}}\approx 110\text{ MeV}$), and $V_{\text{CKM}}^{(m^{-})}$ the CKM matrix element corresponding to two quarks representing the meson ($V_{us}$ for $K^{-}$ and $V_{ud}$ for $\pi^{-}$);
\begin{equation}
g_{\rho^{0}\gamma} = \frac{em_{\rho}^{2}}{\sqrt{3}} \approx 0.135\text{ GeV}^{2}
\end{equation}

The matrix elements for the processes $m^{0}\to \gamma+N+\nu$, where $m^{0} = \pi^{0}/\eta$, are
\begin{multline}
    \mathcal{M}_{\pi^{0}/\eta \to \gamma+N+\nu} = d\cdot 2\cdot\frac{g_{m^{0}\gamma\gamma}}{2}\epsilon^{\mu\kappa\alpha\beta}F_{\mu\kappa}(p_{\gamma})\times \\ \times \frac{1}{p_{(N\nu)}^{2}}\left[p_{(N\nu),\alpha}p_{(N\nu),\rho}g_{\beta\sigma}+p_{(N\nu),\beta}p_{(N\nu),\sigma}g_{\alpha\rho} -p_{(N\nu),\alpha}p_{(N\nu),\sigma}g_{\beta\rho}-p_{(N\nu),\beta}p_{(N\nu),\rho}g_{\alpha\sigma}\right]\times \bar{u}_{L}(p_{\nu})\sigma_{\rho\sigma}u(p_{N})
\end{multline}
Here, $p_{(N\nu),\alpha} = (p_{N}+p_{\nu})_{\alpha}$, and $F_{\mu\nu}(p_{\gamma}) = p_{\gamma,\mu}\epsilon_{\nu}(p_{\gamma})-p_{\gamma,\nu}\epsilon_{\mu}(p_{\gamma})$, with $\epsilon$ being the polarization vector of the photon.

The matrix element for the process $m^{-}\to l^{-} \gamma N$ is
\begin{equation}
    \mathcal{M}_{m^{-} \to l \gamma N} = f_{m^{-}l\nu} = d \cdot F_{\mu\nu}(p_{\gamma})p_{m^{-},\sigma}\bar{u}(p_{l})\gamma^{\sigma}(1-\gamma_{5})D_{(\nu)}(p_{\gamma}+p_{N})\frac{1+\gamma_{5}}{2}\sigma^{\mu\nu}v(p_{N})
\end{equation}
Finally, the matrix element of the decay $\rho^{0}\to N+\nu$ is
\begin{equation}
    \mathcal{M}_{\rho^{0}\to N+\nu} = g_{\rho^{0}\gamma}\epsilon_{\alpha}(p_{\rho})\cdot d\cdot \frac{g_{\alpha\gamma}p_{\rho,\beta}-g_{\alpha\beta}p_{\rho,\gamma}}{p_{\rho}^{2}}\bar{u}_{L}(p_{\nu})\sigma^{\beta\gamma}v(p_{N})
\end{equation}
The decay widths are computed as~\cite{ParticleDataGroup:2020ssz}
\begin{equation}
    \Gamma_{X\to YZ} = \frac{|M|^{2}}{8\pi}\frac{|\mathbf{p}_{Y}|}{m_{X}^{2}}
\end{equation}
for 2-body decays, where $\mathbf{p}_{Y}$ is the momentum of the $Y$ particle in the rest frame of decaying $X$, and
\begin{equation}
    \Gamma_{X\to YZU} = \int dm_{YZ}dm_{ZU} \frac{|M|^{2}}{256\pi^{3}m_{X}^{3}},
\end{equation}
where $m_{ab} = (p_{a}+p_{b})^{2}$.
In the limit of $m_{N,l}\ll m_{\text{meson}}$, the decay widths are
\begin{equation}
    \Gamma_{\rho^{0}\to N\nu} \approx \frac{\alpha_{\text{EM}}d^{2}m_{\rho^{0}}^{3}}{36\pi}, \quad \Gamma_{m^{-}\to \gamma N l^{-}} \approx \frac{d^{2}f_{m^{-}}^{2}G_{F}^{2}m_{m^{-}}^{5}|V_{\text{CKM}}^{(m^{-})}|^{2}}{192\pi^{3}}, \quad  \Gamma_{m^{0}\to N\gamma \nu} \approx \frac{d^{2}g_{m^{0}\gamma\gamma}^{2}m_{m^{0}}^{5}}{96\pi^{3}} \,.
\end{equation}

\begin{figure}[t]
    \centering
    \includegraphics[width=0.7\textwidth]{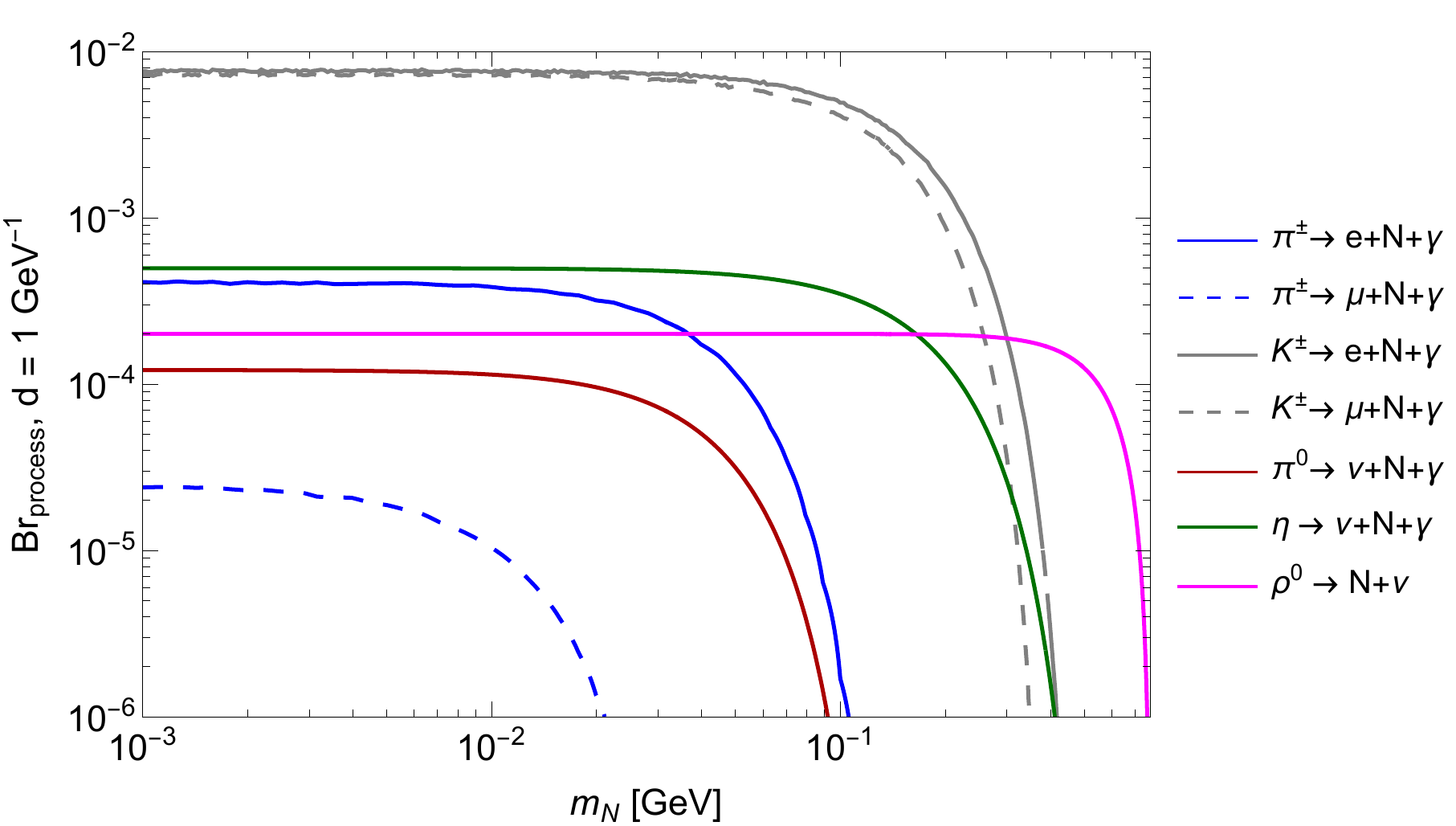}
    \caption{Branching ratios of the HNL production by decays of mesons via the dipole portal (the processes~\eqref{eq:dipole-production-mesons-1},~\eqref{eq:dipole-production-mesons-2}).}
    \label{fig:meson-decays-branching-ratios}
\end{figure}

The resulting branching ratios under the assumption $|d|= 1\text{ GeV}^{-1}$  are shown in Fig.~\ref{fig:meson-decays-branching-ratios}, where we use the SM decay widths of the mesons from~\cite{ParticleDataGroup:2020ssz}. 
The largest branching ratio is for the decay $K^{-}\to N+\gamma+l^{-}$. It is larger than the ratio for the decay $\pi^{-}\to N+\gamma+l^{-}$ due to the scaling $\Gamma_{m^{-}\to N+\gamma+l^{-}} \propto m_{m^{-}}^{5}$. Indeed, the SM decay widths of the charged mesons scale instead according to $\Gamma_{m^{-},\text{SM}} \propto m_{m^{-}}^{3}$. Therefore, we have $\text{Br}_{m^{-}\to N+\gamma+l^{-}} \propto m_{m^{-}}^{2}$.

To estimate the flux of HNLs from the mesons in the direction of the DUNE ND, we have followed the same procedure as described in Appendix~\ref{app:neutrino-meson-flux}.

%%%%%%%%%%%%%%%%%%%%%%%%%%%%%%%%%%%%%%%%%%%%%%%%%%%%%%%%%%%%%%%%%%%%%%%%%
\subsubsection{Neutrino up-scattering -- inside}
\label{app:dipole-portal-production-scatterings}
Neutrino up-scattering is a process
$\nu +T \to N + X$, where $T$ is a target particle, and $X$ denotes an arbitrary final state. In this study, we consider the following processes:
\begin{equation}
    T = e, X = e; \quad T = n/p, X = n/p; \quad T = \text{Ar},X = \text{Ar}; \quad T = n/p, X = \text{hadrons} \,. 
\end{equation}
The first three sets correspond to quasi-elastic (QE) scattering off electrons, nucleons, and Ar nuclei. The last set corresponds to deep inelastic neutrino up-scattering (DIS). 
\begin{figure*}[!t]
    \centering
    \includegraphics[width=0.5\textwidth]{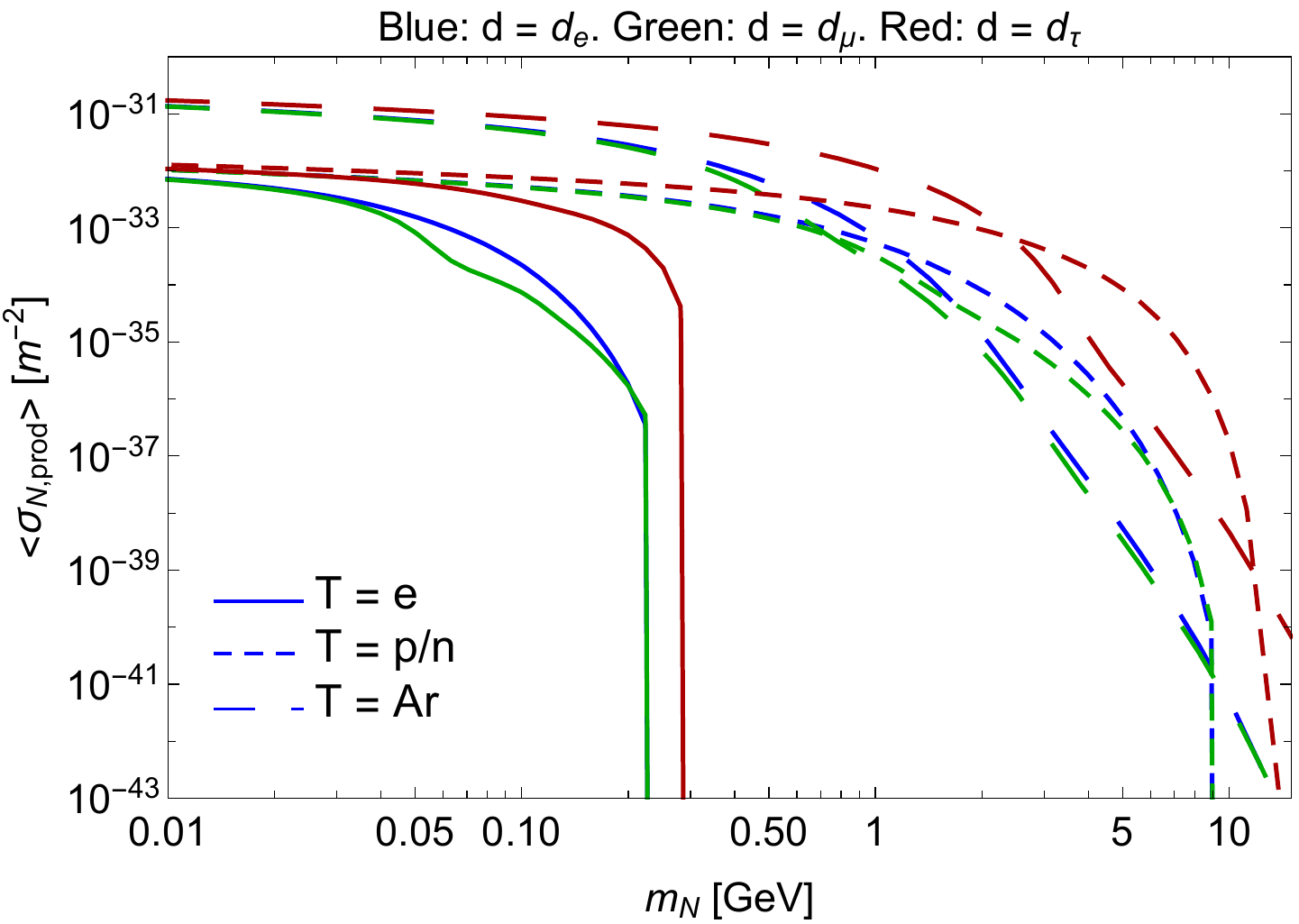}\\\includegraphics[width=0.5\textwidth]{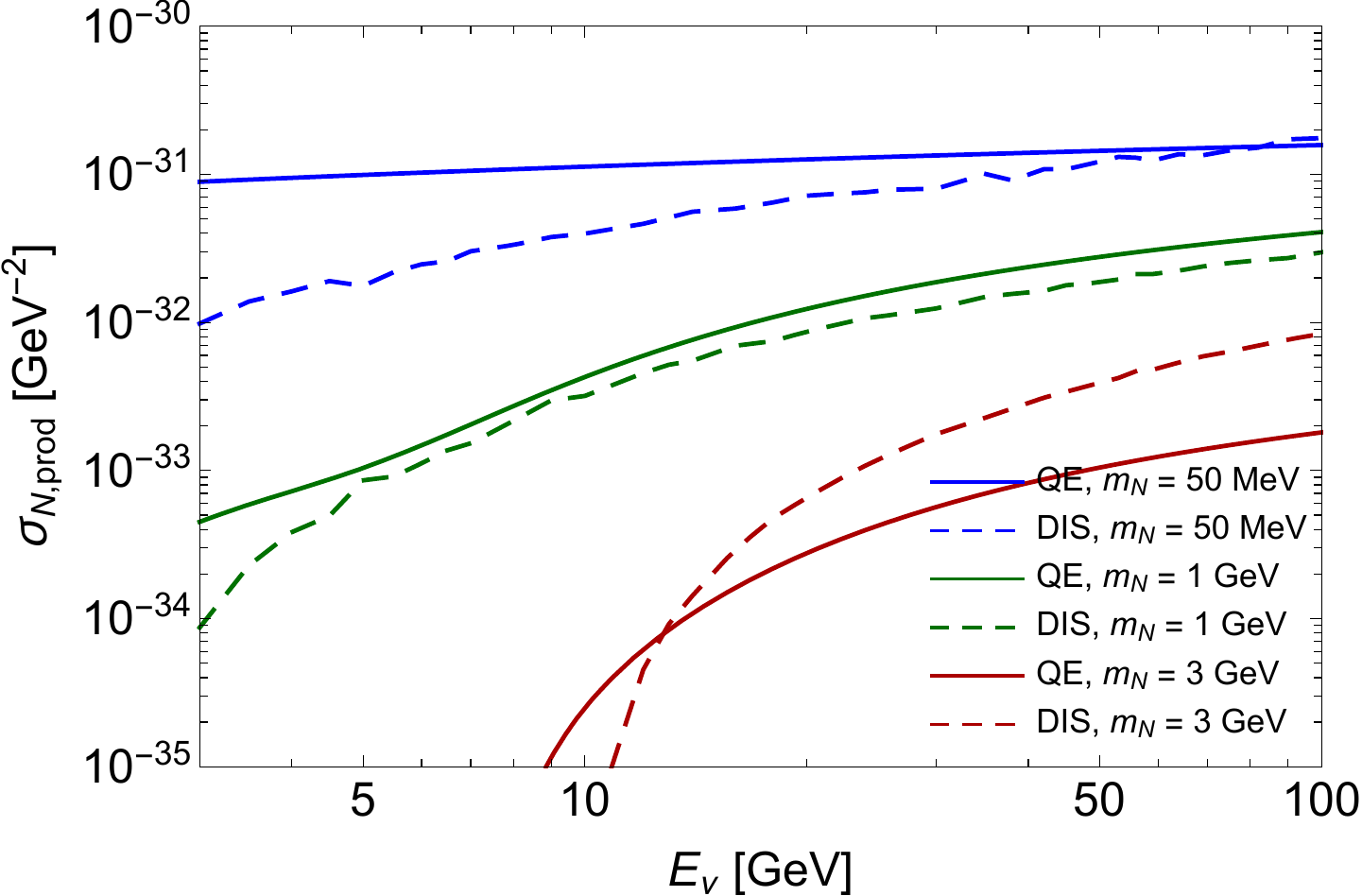}~
    \includegraphics[width=0.5\textwidth]{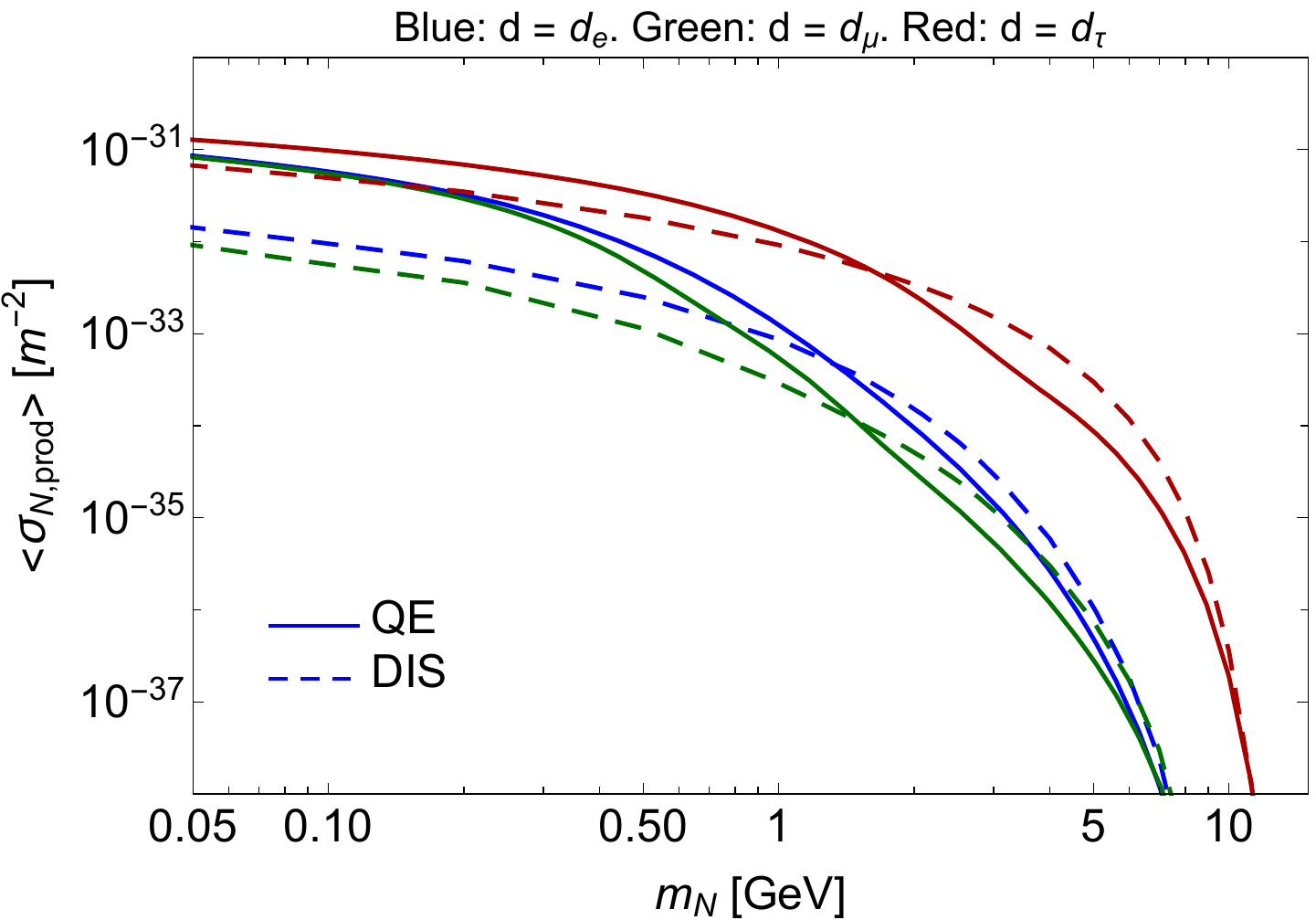}
    \caption{Production cross sections for HNLs at the DUNE ND (per nucleon) for the neutrino up-scattering channel. \textit{Top panel}: the QE cross sections averaged over neutrino energies (Eq.~\eqref{eq:dipole-cross-sections-averaged}) for electron (solid), nucleon (short-dashed) and Ar nucleus (long-dashed) targets. Different colors correspond to $d = d_{e},d_{\mu},d_{\tau}$, where $d_{\alpha} = 1\text{ GeV}^{-1}$. \textit{Bottom left panel}: neutrino energy dependence of the total elastic (solid) and DIS (dashed) cross sections for different HNL masses before averaging. \textit{Bottom right panel}: Mass dependence of the total averaged elastic and DIS cross sections.}
    \label{fig:dipole-portal-up-scattering-cross-sections}
\end{figure*}

The QE differential cross sections $d\sigma_{\text{N,prod}}^{T}(E_{\nu},Q^{2})/dQ^{2}$, with $Q^{2}$ being the modulus-squared of the momentum transferred to the target, have been previously computed, see e.g.~\cite{Schwetz:2020xra}. For the estimates of the number of events, it is useful to compute the cross sections averaged over neutrino energies at the DUNE ND,
\begin{equation}
   \langle \sigma_{\text{N,prod}}^{T}\rangle \equiv \int dE_{\nu} dQ^{2}\ f_{\nu}(E_{\nu}) \frac{d\sigma_{\text{N,prod}}^{T}(E_{\nu},Q^{2})}{dQ^{2}},
   \label{eq:dipole-cross-sections-averaged}
\end{equation}
where $f_{\nu}(E_{\nu})$ is the neutrino distribution function:
\begin{equation}
f_{\nu}(E_{\nu}) \equiv \frac{\Phi_{\nu}(E_{\nu})}{\int dE_{\nu}\ \Phi_{\nu}(E_{\nu})}
\end{equation}
The behavior of these cross sections as a function of the HNL mass and for different target particles is shown in Fig.~\ref{fig:dipole-portal-up-scattering-cross-sections}, (top panel). At small HNL masses, the production from scattering off nuclei dominates, which is because of the enhancement of the cross section (per nucleon) by the factor $Z^{2}/A$. However, the nuclear EM form factor exponentially suppresses large momentum transfers (and hence the production of heavy HNLs), and at masses around $m_{N}\simeq 1\text{ GeV}$, the scattering off nucleons starts to dominate. At mass $m_{N}\simeq 8\text{ GeV}$ (12 GeV) for the production from $\nu_{e/\mu}$ ($\nu_{\tau}$ at ND), the production via scattering off nucleons reaches the kinematic threshold and instantly turns off. The only remaining production channel is the scattering off nuclei, although it is exponentially suppressed.

To calculate the DIS cross section, we have first implemented the model of the dipole portal in MadGraph5~\cite{Alwall:2014hca} with the help of FeynRules~\cite{Alloul:2013bka, Christensen:2008py}. Then, we generated the process 
\begin{equation}
\nu_{\mu}+ p \to N+ \text{jet},
\end{equation} 
with further interfacing to pythia8 for subsequent showering/hadronization. As a cross-check of the model implementation, we have reproduced the result of the analytic formula from~\cite{Schwetz:2020xra} for the production of the HNL in the $\nu$ up-scattering off electrons. As a cross-check of showering, we have reproduced the DIS neutrino cross section $\nu + n \to p + X$ from~\cite{Lalakulich:2013tca} within the systematic uncertainty for the neutrino energy range $1\text{ GeV} < E_{\nu}<40\text{ GeV}$.

The comparison of the DIS cross section with the QE up-scattering cross sections for various HNL masses and neutrino energies, as well as the cross sections averaged over neutrino energies, are shown in Fig.~\ref{fig:dipole-portal-up-scattering-cross-sections}. Before averaging over neutrino energies, the DIS cross section dominates for large HNL masses and, simultaneously, if neutrino energies are large enough to allow momenta transfers $Q^{2}\gg 1\text{ GeV}^{2}$ to the nucleons (bottom left panel). However, the neutrino high-energy tail is strongly suppressed, and after averaging over neutrino energies, this suppression compensates for the improvement, see Fig.~\ref{fig:dipole-portal-up-scattering-cross-sections} (bottom right).

%%%%%%%%%%%%%%%%%%%%%%%%%%%%%%%%%%%%%%%%%%%%%%%%%%%%%%%%%%%%%%%%%%%%
\subsubsection{Neutrino up-scattering -- outside}
\label{app:dipole-portal-production-scatterings-outside}
In addition to the production by up-scatterings inside the detector, HNLs may be produced outside, when neutrinos scatter off the crust separating the decay pipe and the ND, see Fig.~1.2 in Ref.~\cite{DUNE:2020lwj}.

For simplicity, we will drop the contribution of DIS, and estimate the outside production via QE processes. In this way, our estimate is conservative.

\begin{figure}
    \centering
    \includegraphics[width=0.5\textwidth]{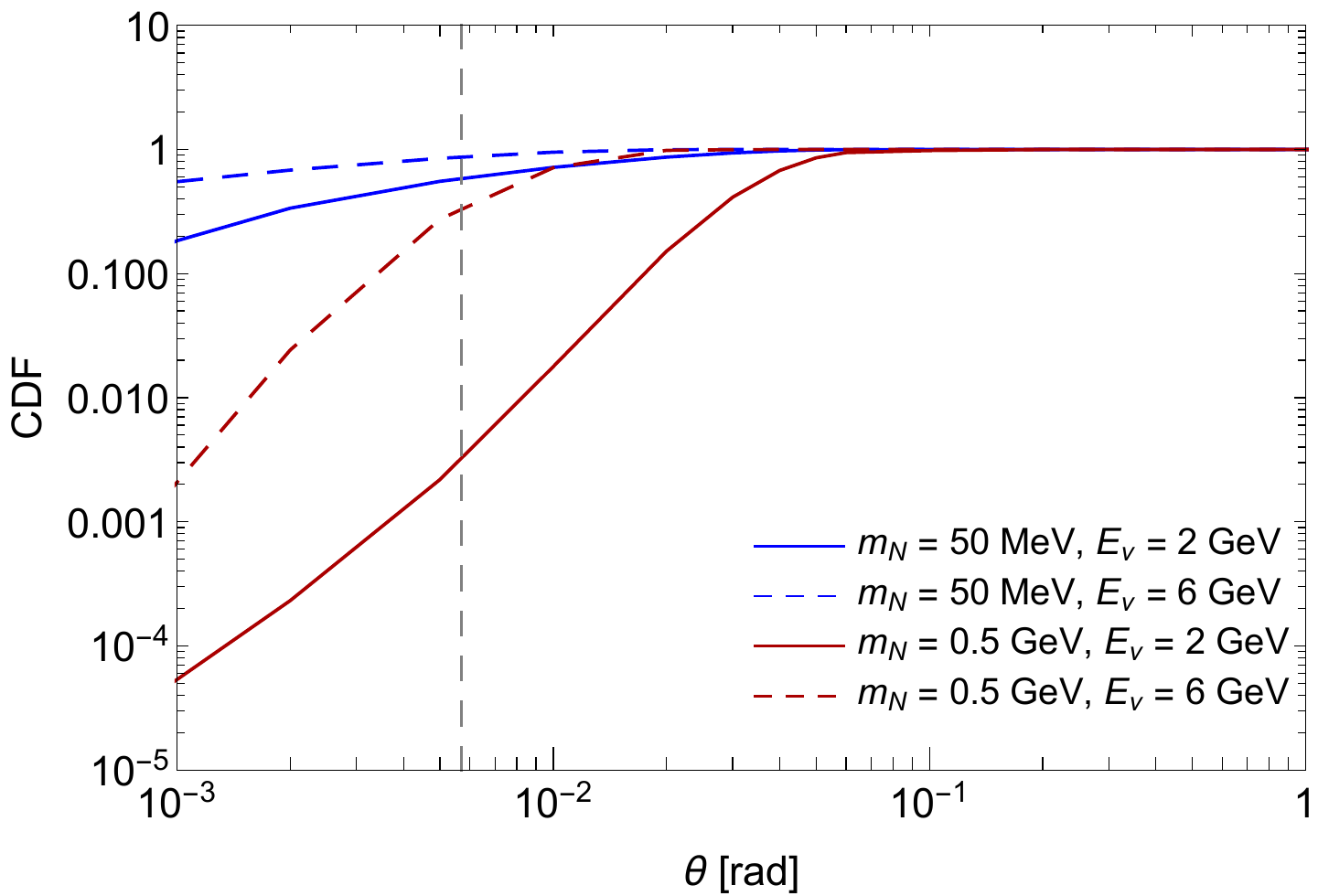}\\ \includegraphics[width=0.45\textwidth]{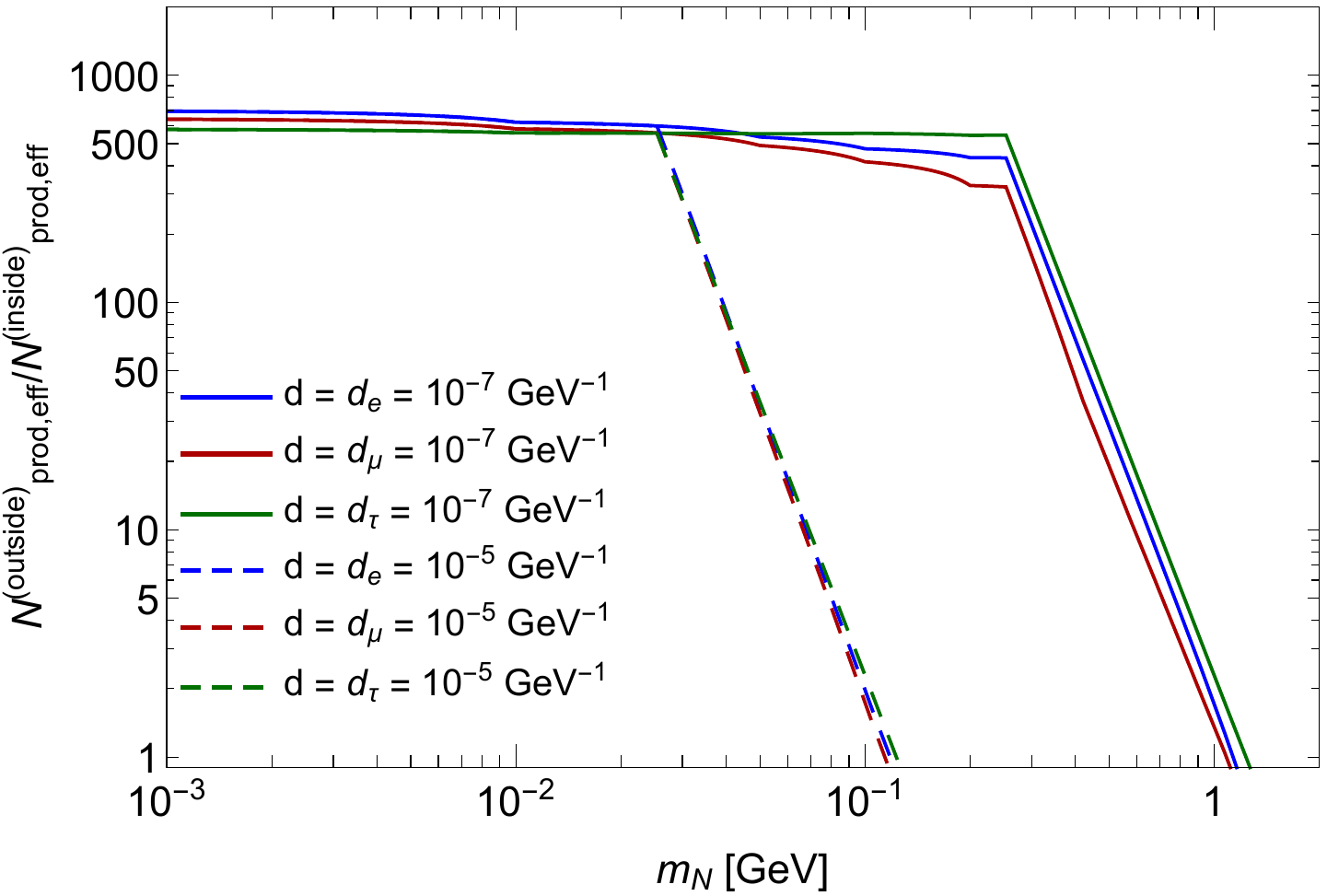}~\includegraphics[width=0.45\textwidth]{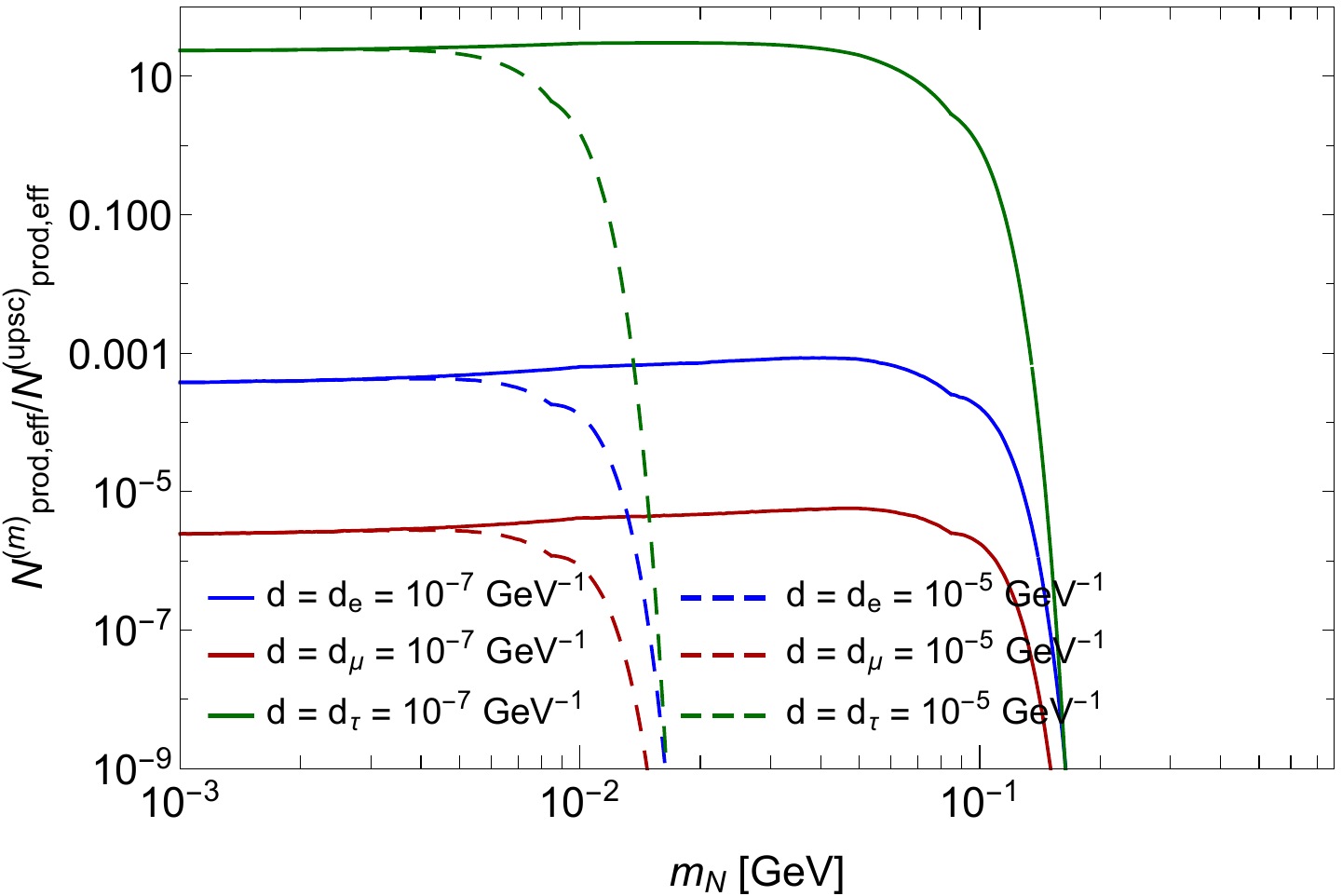}
    \caption{\textit{Top panel}: cumulative distribution function in the angle, $\text{CDF}(\theta) = \frac{1}{\sigma_{\text{N prod}}} \int \limits_{0}^{\theta} d\theta \ \frac{d\sigma_{\text{N prod}}}{d\theta}$, between a neutrino and an HNL produced by up-scatterings off nuclei, shown for several values of the HNL mass and neutrino energy. The vertical dashed line corresponds to the angle $\Delta \theta \approx \sqrt{S_{\text{transverse}}}/(2\times 304\text{ m})\approx 0.005 \text{ rad}$ covered by ND as seen from the end of the decay pipe~\cite{DUNE:2020lwj}. \textit{Bottom left panel}: the behavior of the ratio of the effective number of HNLs produced via outside and inside up-scatterings and pointing to ND, Eq.~\eqref{eq:Nprod-up-scattering-inside-outside}, as a function of the HNL mass and for various choices of $d$. \textit{Bottom right panel}: the ratio of the effective number of HNLs produced by decays of mesons and traveling in the direction of DUNE ND, and neutrino up-scatterings outside DUNE ND (Eq.~\eqref{eq:Nprod-meson-up-scattering}), as a function of the HNL mass and for various choices of $d$.}
    \label{fig:up-scattering-inside-outside}
\end{figure}

A simple estimate of the flux of HNLs produced by the outside up-scatterings may be obtained under the assumption that HNLs travel in the same direction as neutrinos. However, for neutrino energies $E_{\nu} \lesssim 5\text{ GeV}$ that dominate the flux, this is not the case: the typical angle between the neutrino and the produced HNL may significantly exceed the angular size of the detector seen from the end of the decay pipe, see Fig.~\ref{fig:up-scattering-inside-outside}. The broadening of the HNL flux means that in general an HNL pointing to the ND may be produced by a neutrino that does not travel to the ND, and vice versa -- a neutrino that travels to the ND may produce an HNL not reaching the detector.

To estimate the flux of HNLs at the ND, we have followed a procedure similar to the one described in Appendix~\ref{app:neutrino-meson-flux}. Namely, for all neutrinos produced by meson decays (not only the neutrinos in the direction of the ND), we generated a random scattering point located between the end of the decay pipe and the beginning of the ND. Then, we generated the produced HNL angle/energy according to the angle/energy distribution given by the differential up-scattering cross section, and selected only those HNLs that point to the detector. For simplicity, we assume that the crust is made of $^{28}\text{Si}$. Having this dataset, we have computed the total number of HNLs traveling to ND, as well as the distribution function $f^{(\text{upsc})}(l_{N},E_{N})$.

The calculation of the flux of HNLs produced by outside up-scatterings at the FD is technically more complicated, as it requires the flux of neutrinos flying by very small angles $\theta \sim 10^{-6}-10^{-4}\text{ rad}$. To calculate such flux, we have computed the distribution of muon neutrinos in angles and energies semi-analytically, using the approach of Ref.~\cite{Boiarska:2019vid} from Appendix B. We have verified the obtained distribution by comparing the fluxes of neutrinos at ND and FD with the results of the DUNE simulations and finding an excellent agreement~\cite{DUNEfluxes}. Having the neutrino fluxes, we have followed the same procedure as for ND.

%%%%%%%%%%%%%%%%%%%%%%%%%%%%%%%%%%%%%%%%%%%%%%%%%%%%%%%%%%%%%%%%%%%%%%%
\subsubsection{Comparison of different production channels}
\label{app:dipole-portal-production-comparison}
Before calculating the sensitivity, it would be useful to understand which production channel dominates the sensitivity. In this sub-section, we make the comparison between the fluxes of HNLs from up-scatterings and decays of mesons using simple estimates. 

\paragraph{Outside and inside up-scattering.} Let us first compare the number of HNLs produced by up-scattering inside and outside the ND. The comparison is non-trivial, since not all the HNLs produced outside may be able to reach the detector volume: it depends on their decay length $l_{\text{N,\text{decay}}} = c\tau_{N}p_{N}/m_{N}$, where $\tau_{N} = \Gamma_{\text{decay,N}}$ is the HNL lifetime (see Sec.~\ref{sec:dipole-portal-DUNE}). Therefore, instead of the total number of the produced HNLs we consider the (effective) number of produced HNLs:
\begin{equation}
    N^{\text{(upsc)}}_{\text{prod,eff}}= N_{\nu,\text{tot}} \times \sum_{T=\text{nucleus,n/p}}\epsilon_{\text{geom}}^{(T)}\times\langle\sigma_{\text{N,prod}}^{(T)}\rangle\times n_{\text{nucl}} \times L_{\text{scatt,eff}}
    \label{eq:Nprod-up-scattering-inside-outside}
\end{equation}
Here, $N_{\nu,\text{tot}}$ is the total number of neutrinos produced at DUNE. $\epsilon_{\text{geom}}$ is the fraction of neutrinos producing HNLs that travel to the ND; for the inside scatterings, it is just the fraction of neutrinos flying to the ND. $\langle\sigma_{\text{N,prod}}\rangle$ is the mean scattering cross section (per nucleon) of these neutrinos; $n_{\text{nucl}}$ is the nucleon number density; $L_{\text{scatt,eff}}$ is the length available for scatterings, accounting that HNLs have to survive long enough and decay inside the detector.

For the outside events, we assume the crust number density $n_{\text{nucl}}\approx 1.4\cdot 10^{30}\text{ m}^{-3}$~\cite{ParticleDataGroup:2020ssz}, and 
\begin{equation}
L_{\text{scatt,eff}} =\text{min}[L_{\text{crust}},l_{\text{decay,N}}(m_{N},d)],
\end{equation}
where $L_{\text{crust}} = 304\text{ m}$ or $\simeq 1300\text{ km}$ is the distance between the end of the decay pipe and the beginning of the ND/FD~\cite{DUNE:2020lwj}. It accounts for the fact that only neutrinos scattering in locations closer than $\simeq l_{\text{decay,N}}$ effectively contribute to the flux of HNLs that can reach the detector. We assume $p_{N} = 10\text{ GeV}$ as the characteristic momentum of short-lived HNLs. 

For the inside events, we take $n_{\text{nucl}} \approx 8.4\cdot 10^{29}\text{ m}^{-3}$ (see Table~\ref{tab:DUNE-parameters}) and $L_{\text{scatt,eff}} = L_{\text{det}}/2 = 2\text{ m}$ (the origin of the factor of 1/2 is explained around Eq.~\eqref{eq:Pdecay-up-scattering-limit}). 

The difference in $\langle\sigma_{\text{N,prod}}^{\text{(i)}}\rangle$ at the ND/FD and at crust is $\mathcal{O}(1)$, given by somewhat different neutrino distributions and different charge/mass number of nuclei. Therefore, the ratio of the number of produced HNLs outside and inside the detector is
\begin{equation}
    \frac{N_{\text{prod,eff}}^{\text{(outside)}}}{N_{\text{prod,eff}}^{\text{(inside)}}} \simeq \frac{\epsilon_{\text{geom}}^{\text{(outside)}}}{\epsilon_{\text{geom}}^{\text{(inside)}}}\times \frac{n_{\text{nucl}}^{\text{outside}}}{n_{\text{nucl}}^{\text{inside}}} \times \frac{L_{\text{scatt,eff}}^{\text{outside}}}{L_{\text{scatt,eff}}^{\text{inside}}} \simeq 
    \begin{cases} 
    250\frac{\epsilon_{\text{geom}}^{\text{(outside)}}}{\epsilon_{\text{geom}}^{\text{(inside)}}} \qquad \text{(ND)}, \\ 
    1.8\cdot 10^{4}\frac{\epsilon_{\text{geom}}^{\text{(outside)}}}{\epsilon_{\text{geom}}^{\text{(inside)}}} \quad \text{(FD)}. \end{cases}
    \label{eq:ratio-up-scattering-outside-inside}
\end{equation}
The behavior of the ratio~\eqref{eq:ratio-up-scattering-outside-inside} for the case of ND is shown in Fig.~\ref{fig:up-scattering-inside-outside}. We see that the geometric acceptance provides a $\mathcal{O}(1)$ correction. For the behavior of this ratio at the FD, the situation is qualitatively similar. 

The ratios~\eqref{eq:ratio-up-scattering-outside-inside} contradict the results of the paper~\cite{Schwetz:2020xra}, according to which the outside up-scattering does not contribute to the sensitivity of the ND at all. We note that the outside fluxes have been calculated differently: using the pre-computed off-axis neutrino flux in the GLoBES format and the analytic formula in~\cite{Schwetz:2020xra}, whereas in this work we re-simulate the neutrino flux and the HNL production event-by-event. It is worth mentioning that the GLoBES files include the high-energy neutrino tail, and hence the difference cannot be attributed to the latter. We have not found an explicit reason for the discrepancy between the results, but there is a simple argument showing that the number of events from outside up-scattering at the ND is at least comparable with the inside flux: let us consider a volume of the crust material having the volume equal to the volume of the ND, and placed just in front of it. Neutrinos up-scatterings inside this volume would produce an HNL flux being a factor of $(L_{\text{det}}/L_{\text{scatt,eff}})\times (n_{\text{nucl}}^{\text{crust}}/n_{\text{nucl}}^{\text{ND}}) \simeq 3$ larger than the inside flux. Given the location of the considered crust volume, all of these HNLs would reach the ND and provide a decay signal, in the relevant limit where the decay length of the HNLs is large compared to $L_{\rm det}$. 

\paragraph{Production from mesons and outside up-scatterings.} The HNL production from mesons is relevant only at ND. The reason is that all mesons decay before the end of the decay pipe, and to be able to reach the FD, the produced HNLs have to travel the distance of $\mathcal{O}(1300\text{ km})$. Since the HNL decay length $c\tau_{N}\gamma_{N}\propto d^{-2}$, this is possible only if $d$ is very small, which a priori suppresses the flux of the HNLs.

Let us now compare the amount of HNLs produced by up-scattering, and those produced by mesons and flying in the direction of the ND. Again, we should compare an effective number of particles, defined by 
\begin{equation}
    N_{\text{prod,eff}} = 
    \begin{cases}
    \sum_{m}N_{\text{m,tot}}\times \text{Br}(m\to N)\times \epsilon_{\text{geom}}^{(m)}\times \text{exp}[-L_{\text{min,eff}}/l_{\text{decay,N}}] \quad \text{(mesons)},\\ 
    N_{\nu}\times \sum_{T}\langle \sigma_{N,\text{prod}}^{(T)}\rangle \times n_{\text{nucl}}\times L_{\text{scatt,eff}} \qquad\qquad \text{(up-scattering)}.
    \end{cases}
    \label{eq:Nprod-meson-up-scattering}
\end{equation}
Here, $N_{\text{m,tot}}$ is the total number of mesons $m$ produced at DUNE, $\epsilon_{\text{geom}}$ is the fraction of produced HNLs that travel to the detector and $\text{exp}[-L_{\text{min,eff}}/l_{\text{decay,N}}]$ accounts for the exponential suppression of the flux of short-lived HNLs produced by decays of mesons. In particular, for $d = d_{e,\mu}$, HNLs are produced by long-lived mesons, and we fix $L_{\text{min,eff}}\approx 221\text{ m}$, which is the end of the decay pipe \cite{DUNE:2020lwj}. For $d = d_{\tau}$, where the production comes from promptly decaying mesons, we fix $L_{\text{min,eff}} = L_{\text{to det}} = 574\text{ m}$. 

The ratio of the produced HNLs resulting from Eq.~\eqref{eq:Nprod-meson-up-scattering} is shown in Fig.~\ref{fig:up-scattering-inside-outside}. From the figure, we see that except for the production via $d_{\tau}$, for which the $\nu_{\tau}$ flux is strongly suppressed, the production from neutrino up-scattering dominates. The reason is the relative suppression of the probability to produce HNLs by decays of mesons ($L_{\text{crust}} = 304\text{ m}$): 
\begin{equation}
\frac{\text{Br}(\text{m}\to N)}{P_{\text{scattering}}} \approx  \frac{\text{Br}(\text{m}\to N)}{ \langle\sigma_{N,\text{prod}}^{(T)}\rangle\times n_{\text{nucl}}\times L_{\text{crust}}} \lesssim 10^{-4}-10^{-2},
\end{equation}
in dependence on the meson which may decay into an HNL, see Figures~\ref{fig:dipole-portal-up-scattering-cross-sections},~\ref{fig:dipole-decay-br-ratios}.

\subsection{Number of events at DUNE}
\label{app:dipole-portal-Nevents}

The number of HNL events for the case of non-zero coupling $d_{\alpha}$ consists of two parts:
\begin{equation}
   N_{\text{events}} =  N_{\text{events}}^{\text{up-scattering}}+N_{\text{events}}^{\text{mesons}},
\end{equation}
which corresponds to the contribution from neutrino up-scattering and decays of mesons. The number of events from inside up-scatterings per one year is
\begin{equation}
    N_{\text{events}}^{\text{upsc, in}} \approx 
    N_{\nu,\text{detector}}\times \sum_{T = e,N,\text{Ar},\text{DIS}} \int dE_{N} \ \frac{d\langle\sigma_{\text{N prod}}^{(T)}\rangle_{E_{\nu}}}{dE_{N}} n^{\text{det}}_{\text{nucl}} \times \langle P_{\text{decay}}(E_{N})\cdot L\rangle \times \text{Br}(N\to \text{channel})
    \label{eq:dipole-portal-nevents-up-scattering}
\end{equation}
Here, $N_{\nu,\text{detector}} = S_{\text{transverse}}\int \Phi_{\nu}(E_{\nu})dE_{\nu}$ is the total number of neutrinos flying in the direction of the detector, with $S_{\text{transverse}}$ being the detector's transverse area (see Table~\ref{tab:DUNE-parameters}; note that for the FD we include four modules, which in total have twice larger transverse area than one module). $d\langle\sigma_{\text{N prod}}^{(T)}\rangle_{E_{\nu}}/dE_{N}$ is the differential production cross section (per nucleon) averaged over neutrino energies. $n^{\text{det}}_{\text{nucl}}$ is the nucleon number density (which corresponds to the liquid agron for both ND and FD). $\langle P_{\text{decay}}(E_{N})\cdot L\rangle$ is the HNL decay probability averaged over length of the detector $L$ available for neutrino scatterings:
\begin{equation}
    \langle P_{\text{decay}}(E_{N})\cdot L\rangle = \int \limits_{0}^{L_{\text{det}}-L_{\text{min}}} dL \ \frac{L}{l_{N,\text{decay}}}\times \exp\left[-(L_{\text{det}}-L)/l_{N,\text{decay}}\right], \quad l_{N,\text{decay}} = c\tau_{N}p_{N}/m_{N}
    \label{eq:Pdecay-up-scattering}
\end{equation}
where the factor $L$ comes from the production probability in the neutrino up-scatterings, $P_{\text{prod}} = \sigma n_{\text{nucl}} L$, and $\tau_{N}^{-1} = \frac{d^{2}m_{N}^{3}}{4\pi}$. The parameter $L_{\text{min}}$ is either 0 or $\Delta l_{\text{DUNE}} = 1\text{ cm}$, depending on the considered signature (displaced or prompt decays). In the limit $l_{\text{N,\text{decay}}}\gg L_{\text{min}}$, this expression simplifies to
\begin{equation}
    \langle P_{\text{decay}}(E_{N})\cdot L\rangle \approx \frac{L_{\text{det}}^{2}}{2l_{\text{decay,N}}}
    \label{eq:Pdecay-up-scattering-limit}
\end{equation}
Finally, $\text{Br}(N\to \text{channel})$ is the branching ratio of the HNL decay into the given particle state, which is $\nu \gamma$, $\nu ee$, $\nu\mu\mu$ depending on the signature.
The approximation made in~\eqref{eq:dipole-portal-nevents-up-scattering} is that both neutrinos and HNLs fly along the beam axis. This is justified since the DUNE ND and FD cover a small solid angle and are located on axis.

For the number of events from the outside up-scatterings, we have
\begin{equation}
    N_{\text{events}}^{\text{upsc, out}} \approx
    N_{\nu,\text{tot}}\times  \sum_{T = (e,n/p,\text{Si})}\langle\sigma_{\text{N prod}}^{(T)}\rangle \times n_{\text{nucl}}^{\text{crust}}\times \epsilon_{\text{geom}}^{(T)} \times \int \limits_{m_{N}} dE_{N} \ \int \limits_{0}^{L_{\text{to det}}}dl_{N} f^{(T)}(E_{N},l_{N})P_{\text{decay}}(E_{N},l_{N}),
\end{equation}
$P_{\text{decay}}(E_{N},l_{N})$ is the decay probability for an HNL with the energy $E_{N}$ produced at a point located at the distance $l_{N}$ from the beginning of the detector,
\begin{equation}
    P_{\text{decay}}(E_{N},l_{N}) = \exp[-l_{N}/l_{\text{decay,N}}]-\exp[-(l_{N}+L_{\text{det}})/l_{\text{decay,N}}],
\end{equation}
and $f(E_{N},l_{N})$ is the HNL distribution normalized by unity.

Finally, the number of events from the decays of mesons is
\begin{equation}
    N_{\text{events}}^{\text{mesons}} \approx \sum_{m}
    N_{m}\times \text{Br}(m\to N)\times \epsilon_{\text{geom}}^{(m)} \times \int \limits_{m_{N}} dE_{N} \ \int \limits_{0}^{L_{\text{pipe}}}dl_{N} f^{(m)}(E_{N},l_{N})P_{\text{decay}}(E_{N},l_{N})
\end{equation}
Here, $N_{\text{m,tot}}$, $\epsilon_{\text{geom}}^{(m)}$ have been already introduced in Eq.~\eqref{eq:Nprod-meson-up-scattering}. $L_{\text{pipe}}\approx 221\text{ m}$ is the distance from the DUNE target to the end of the decay pipe. For short-lived mesons such as $m = \pi^{0}/\eta/\rho^{0}$, $f^{(m)}(E_{N},l_{N}) \approx f^{(m)}(E_{N})\times \delta(l_{N}-574\text{ m})$.

%%%%%%%%%%%%%%%%%%%%%%%%%%%%%%%%%%%%%%%%%%%%%%%%%%%%%%%%%%
\subsection{The shape of the sensitivity curves}
\label{app:dipole-sensitivity-shape}

The combined sensitivities to inside up-scattering events shown in Fig.~\ref{fig:dipole-portal-sensitivity} behave non-trivially with the HNL mass. Namely at small HNL masses $m_{N}\lesssim 2\text{ GeV}$, the lower bound $d_{\text{lower}}$ decreases, reaches its minimum at some mass $m_{\text{N,peak}}$, and then starts increasing. 
The reason is the following. Schematically, the number of events is $N_{\text{events}} = N_{\text{N,prod}}\times P_{\text{decay}}$. Since $P_{\text{decay}} \leq 1$, the sensitivity cannot cover couplings smaller than defined by the condition $N_{\text{N,prod}}=2$. The number of produced HNLs is $N_{\text{N,prod}} = f(m_{N})\cdot d^{2}$, where $f(m_{N}) \propto \langle\sigma_{\text{N,prod}}\rangle$ is the production cross section. As a result, we may define the minimal coupling $d_{N_{\text{prod}} = 2} = \sqrt{2/f(m_{N})}$. At HNL masses $m_{N}\lesssim 1\text{ GeV}$, $d_{N_{\text{prod}} = 2}$ remains practically constant. The reason is that the cross section weakly depends on $m_{N}$ in this range (Fig.~\ref{fig:dipole-portal-up-scattering-cross-sections}). 

In the domain of small HNL masses and at $d_{N_{\text{prod}} = 2}$, the HNL decay length $l_{\text{N,decay}} = c\tau_{N}\gamma_{N}\propto m_{N}^{-4}d^{-2}$ is parametrically very large, $l_{\text{N,decay}}\gg L_{\text{to det}}$. Therefore, the decay probability is $P_{\text{decay}}\approx L_{\text{det}}/l_{\text{N,decay}} \ll 1$ (remind Eq.~\eqref{eq:Pdecay-up-scattering-limit}). Plugging this expression in the expression for the number of events, we get
\begin{equation}
    N_{\text{events}} \propto f(m_{N})d^{2}\times m_{N}^{4}d^{2} = f(m_{N})m_{N}^{4} \cdot d^{4},
\end{equation}
which is a growing function of the HNL mass. This scaling explains why the lower bound decreases.

However, with the increase of the HNL mass, (i) the HNL decay length at fixed $d$ grows, (ii) $f(m_{N})$ drops, which leads to the increase of $d_{N_{\text{prod}} = 2}$. As a result, $P_{\text{decay}}(d_{N_{\text{prod}} = 2})$ becomes $\mathcal{O}(1)$, and mass dependence of the lower bound becomes to be determined only by the behavior of $f(m_{N})$, which quickly drops at these masses: $d_{\text{lower}} \approx d_{N_{\text{prod}} = 2}$. The position of the minimum of the lower bound, which we denoted as $m_{\text{N,peak}}$, may be estimated as $N_{\text{events}}(m_{N},d_{N_{\text{prod}} = 2}) \simeq 2$.

%%%%%%%%%%%%%%%%%%%%%%%%%%%%%%%%%%%%%%%%%%%%%%%%%
\section{Neutrinophilic scalar portal}
\label{app:neutrinophilic-portal}
%%%%%%%%%%%%%%%%%%%%%%%%%%%%%%%%%%%%%%%%%%%%%%%%%

\subsection{Production}
\label{app:neutrinophilic-production}

\subsubsection{Quasi-elastic production}
Consider the production process
\begin{equation}
    \nu_{\mu}+p \to n + \phi + \mu^{+}
    \label{eq:neutrinophilic-production-quasielastic}
\end{equation}
The matrix element of this process has the form
\begin{equation}
    \mathcal{M} \approx \frac{g_{\phi}G_{F}}{2\sqrt{2}}\cos(\theta_{c})\bar{v}(p_{\nu})D_{\nu}(p_{\nu}-p_{\phi})\gamma^{\mu}P_{L}v(p_{\mu}) \bar{u}(p_{n})\Gamma_{\mu}u(p_{p})
\end{equation}
The $pn$ effective vertex $\Gamma_{\mu}$ is ~\cite{Leitner2005,Zeller:2003ey,LlewellynSmith:1971uhs}
\begin{equation}
\Gamma_{\mu}(p_{p},p_{n}) = \gamma_{\mu}(F_{V1}(q^{2}) -\gamma_{5}F_{A}(q^{2})) - \frac{i}{2m_{p}}\sigma_{\mu\nu}q^{\nu}F_{V2}(q^{2}) - \frac{q_{\mu}}{m_{p}}\gamma_{5}F_{P}(q^{2}),
\label{eq:effective-nucleon-vertex}
\end{equation}
where $q \equiv p_{p}-p_{n}$ is the momentum transferred to the nucleon, and $F$ are form factors. Their explicit expressions are
\begin{equation}
   F_{A}(q^{2}) = \frac{g_{A}}{\left(1+|q^{2}|/m_{A}^{2}\right)^{2}}, \quad F_{P}(q^{2}) = \frac{2m_{p}^{2}}{|q^{2}|+m_{\pi}^{2}}F_{A}(q^{2}),
    \label{eq:form-factors-1}
\end{equation}
\begin{equation}
    F_{V1}(q^{2}) = \frac{G_{E}^{p}(q^{2})-G_{E}^{n}(q^{2})+\tau (G_{M}^{p}(q^{2})-G_{M}^{n}(q^{2}))}{1+\tau},
    \end{equation}
    \begin{equation}
    F_{V2}(q^{2}) = \frac{G_{M}^{p}(q^{2})-G_{M}^{n}(q^{2})- (G_{E}^{p}(q^{2})-G_{E}^{n}(q^{2}))}{1+\tau},
\end{equation}
where $\tau = |q^{2}|/(4m_{p}^{2})$, and $G_{E/P}$ are electric and magnetic form factors. In dipole approximation, they read
\begin{equation}
    G_{E}^{p}(q^{2}) \approx G_{D}(q^{2}), \quad G_{E}^{n}(q^{2}) = -\mu_{n}\frac{a\tau}{1+b\tau}G_{D}(q^{2}),
\end{equation}
\begin{equation}
    G_{M}^{p}(q^{2}) \approx \mu_{p}G_{D}(q^{2}), \quad G_{M}^{n}(q^{2}) = \mu_{n}G_{D}(q^{2}), \quad G_{D}(q^{2}) = \frac{1}{\left(1+\frac{|q^{2}|}{m_{V}^{2}}\right)^{2}},
    \label{eq:form-factors-2}
\end{equation}
with $\mu_{p} = 2.793$, $\mu_{n} = -1.913$ are magnetic moments of the proton and the neutron. The other phenomenological parameters entering the form factors are given in Table~\ref{tab:form-factors-parameters}.
\begin{table}[t]
    \centering
    \begin{tabular}{|c|c|c|c|c|c|c|c|}
        \hline Parameter & $a$ & $b$ & $m_{A}$ & $m_{V}$ & $g_{A}$ \\ \hline
         Value & $0.942$ & $4.61$ & $1.026\text{ GeV}$ & $0.843\text{ GeV}$ & $1.26$ \\ \hline
    \end{tabular}
    \caption{Values of parameters entering the form factors~\eqref{eq:form-factors-1},~\eqref{eq:form-factors-2}.}
    \label{tab:form-factors-parameters}
\end{table}

To calculate the cross section of the process~\eqref{eq:neutrinophilic-production-quasielastic}, we follow~\cite{Byckling:1971vca} for the evaluation of the phase space of $2\to 3$ processes. As a cross-check of the form factors, we have reproduced the energy dependence of the SM quasi-elastic process $\nu_{\mu}+n\to \mu+p$, while in order to verify the $2\to 3$ phase space calculation we have obtained the cross section of the process $\nu_{\mu}+\gamma \to \nu_{\mu}+\mu^{+} + \mu^{-}$ from~\cite{Altmannshofer:2014pba}.

\begin{figure}[t]
    \centering
    \includegraphics[width=0.45\textwidth]{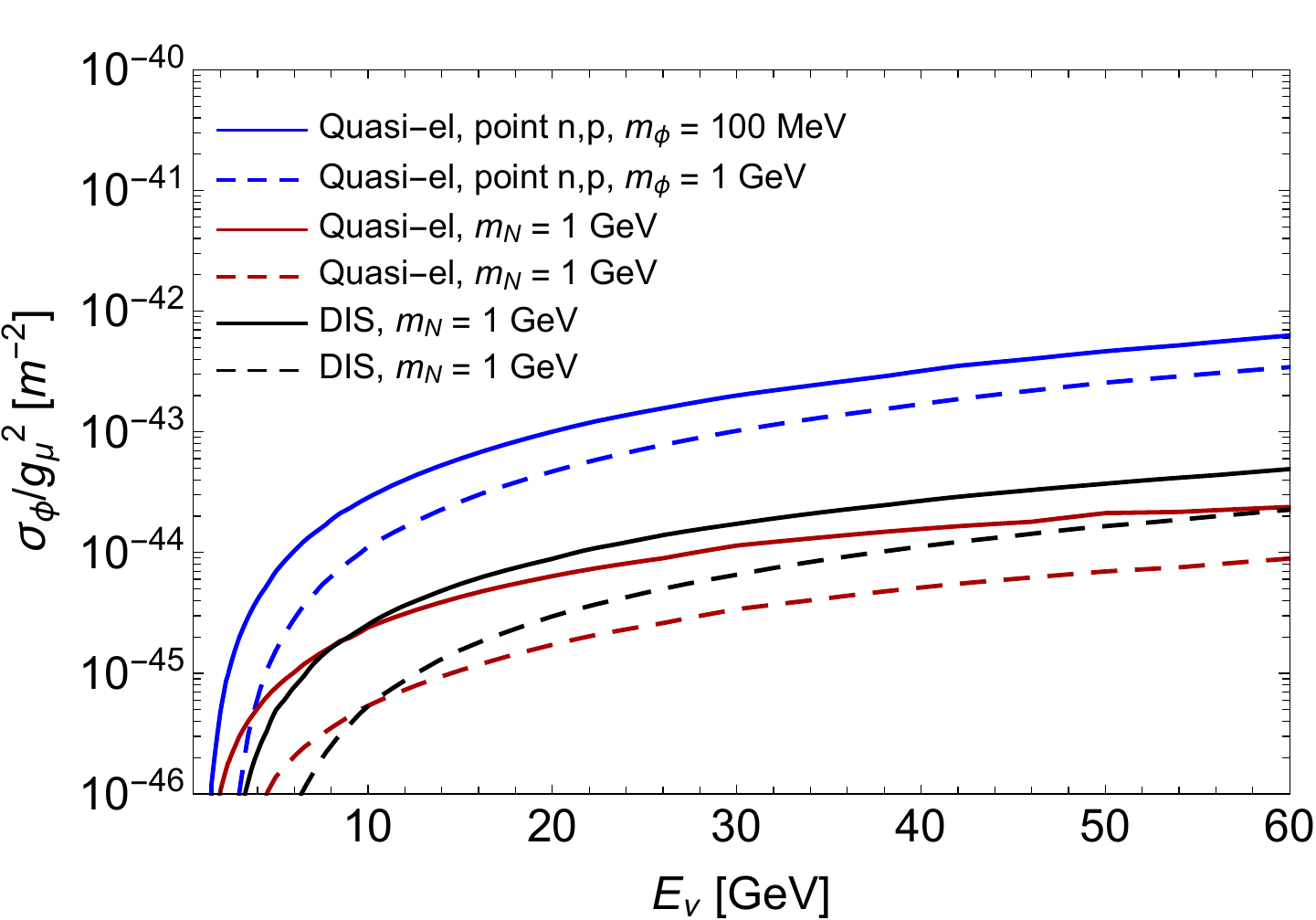}
    \includegraphics[width=0.45\textwidth]{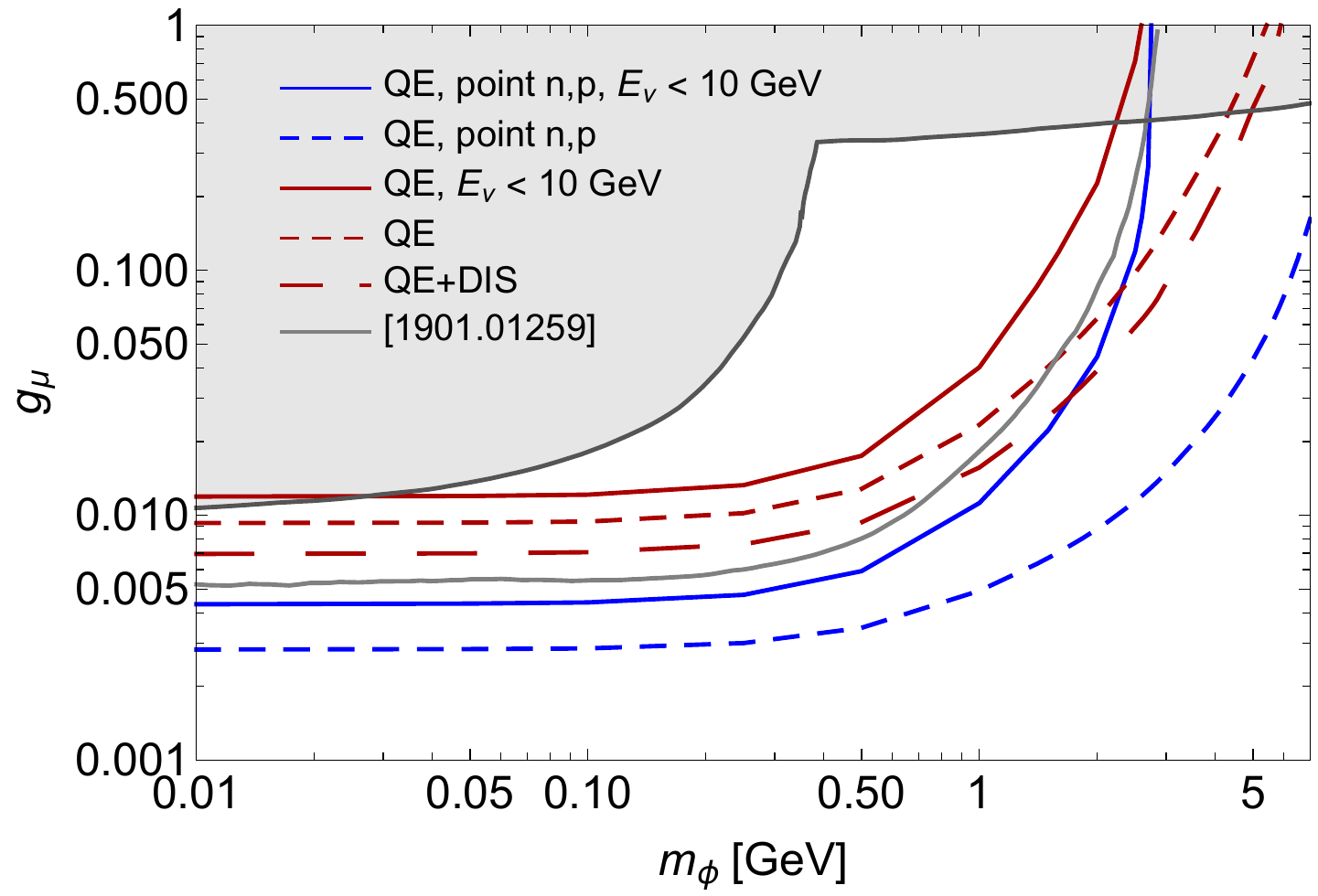}
    \caption{Left panel: the dependence of the cross section of the process~\eqref{eq:neutrinophilic-production-quasielastic} and the DIS production cross section on the neutrino energy for different choices of $\phi$ masses. To illustrate the effect of the form factors entering the effective nucleon vertex~\eqref{eq:effective-nucleon-vertex}, we also show the cross section in the limit~\eqref{eq:point-like-nucleons}.
    Right panel: Impact of QE form factors, the high-energy neutrino tail, and the inclusion of DIS production on the 
    the sensitivity of the DUNE ND to the neutrinophilic scalar portal~\eqref{eq:neutrinophilic-portal}. For comparison we also show the sensitivity obtained in~\cite{Kelly:2019wow}.}
    \label{fig:neutrinophilic-cross-section}
\end{figure}

The neutrino energy dependence of the cross section for the process~\eqref{eq:neutrinophilic-production-quasielastic} is shown in Fig.~\ref{fig:neutrinophilic-cross-section}. For comparison, we also show the cross section with the form factors approximated by their zero momentum limit $q_{\mu}\to 0$:
\begin{equation}
    F_{A}\approx 1.26,\quad F_{V1}\approx 1, \quad F_{V2}=F_{P} = 0\,.
    \label{eq:point-like-nucleons}
\end{equation}

\subsubsection{Deep inelastic scattering production}
The DIS production process is
\begin{equation}
    \nu_{\mu}+p \to \mu^{+}+X+\phi,
    \label{eq:neutrinophilic-production-dis}
\end{equation}
where $X$ denotes any multi-hadron state.
To calculate the DIS cross section, we have implemented the Lagrangian from Eq.~\eqref{eq:neutrinophilic-portal} in MadGraph5 using FeynRules. To check the implementation, we have also included the weak interaction of point-like nucleons~\eqref{eq:point-like-nucleons} and reproduced the quasi-elastic scattering cross section~\eqref{eq:neutrinophilic-quasi-elastic}. The behavior of the DIS cross section with neutrino energy is shown in Fig.~\ref{fig:neutrinophilic-cross-section}. Similarly to the case of the dipole portal, the DIS cross section is smaller than the quasi-elastic cross section at small neutrino energies but becomes larger at high energies.

%%%%%%%%%%%%%%%%%%%%%%%%%%%%%%%%%%%%%%%%%%%%%%%%%%%%%%%%%%%%%%%%
\subsection{Number of events and discussion of sensitivity}
\label{app:neutrinophilic-literature}

The number of events for the neutrinophilic scalar portal at the DUNE ND has the form
\begin{equation}
   N_{\text{events}} =  \sum_{i = \text{quasi-el,DIS}}N_{\nu}\times \sigma_{i}n_{\text{nucl}}L_{\text{det}}, 
   \label{eq:neutrinophilic-Nevents}
\end{equation}
where $n_{\text{nucl}}$ is the number density of nucleons, $L_{\text{det}}$ is the DUNE ND detector length, and $\sigma_{i}$ are the production cross sections for the processes~\eqref{eq:neutrinophilic-production-quasielastic}~\eqref{eq:neutrinophilic-production-dis}, where we require $p_{T,\phi}>0.5\text{ GeV}$. To obtain the sensitivity, we require $N_{\text{events}}>2.3$, corresponding to $90\%$ C.L. in background free regime.

The comparison with~\cite{Kelly:2019wow} under the same assumptions is shown in Fig.~\ref{fig:neutrinophilic-cross-section} (blue solid versus grey curves in the right panel); we find good agreement.
To illustrate the effects of the inclusion of the form factors and of neutrino energy cut, in the figure, we include sensitivity curves under different assumptions about them. We see that the form factors affect not only the domain of large $\phi$ masses but even the domain of small masses $m_{\phi} \ll 1\text{ GeV}$. This is caused by the transverse momentum cut $p_{T,\phi} > 0.5\text{ GeV}$, which sets a lower bound on the momentum transferred to the nucleons. The suppression caused by form factors increases with the growth of the transferred momentum, and therefore the sensitivity worsens. The comparison of the short-dashed and long-dashed red curves in Fig.~\ref{fig:neutrinophilic-cross-section} (right panel) shows the impact of including the DIS production channel.

\end{document}